\documentclass[sigconf,screen]{acmart}

\acmConference[ICSE 2024]{46th International Conference on Software Engineering}{April 2024}{Lisbon, Portugal}

\AtBeginDocument{%
  \providecommand\BibTeX{{%
    \normalfont B\kern-0.5em{\scshape i\kern-0.25em b}\kern-0.8em\TeX}}}

\usepackage{xurl}
\usepackage{multirow}
\usepackage{caption} 
\usepackage{booktabs} 
\usepackage{amsmath}
\usepackage{graphicx}
\usepackage[belowskip=-12pt]{caption}
\usepackage{rotating}
\usepackage{rotating}
\usepackage{tabularx}
\usepackage{subfig}
\usepackage{wrapfig}
\usepackage{listings,chngcntr}
\usepackage{color, colortbl}
\usepackage{balance} 
\usepackage{lscape}
\usepackage{algorithm}
\usepackage[noend]{algpseudocode}
\algnewcommand\algorithmicforeach{\textbf{for each}}
\algdef{S}[FOR]{ForEach}[1]{\algorithmicforeach\ #1\ \algorithmicdo}

\usepackage{xspace}
\usepackage{framed}
\usepackage{syntax}
\usepackage[tikz]{bclogo}
\usepackage{soul}
\usepackage{flushend}
\usepackage{subfig}
\usepackage{pifont}
\newcounter{NumObservations}
\stepcounter{NumObservations}
\definecolor{shadecolor}{rgb}{.9,.9,.9}

\newcommand{\figref}[1]{Fig.~\ref{#1}}

\newcommand{\secref}[1]{\S\ref{#1}}

\newcommand{\keras}{\textit{Keras}\xspace}
\newcommand{\dense}{\texttt{Dense}\xspace}

\newcommand{\wpre}{\textsc{\textit{$wp$}}\xspace}

\newcommand{\deepinfer}{\textit{DeepInfer}\xspace}

\newcommand{\selfchecker}{\textit{SelfChecker}\xspace}

\usepackage{titlecaps}
\newcommand{\twp}{{\scshape (wp)}\xspace}
\newcommand{\twpalpha}{{\scshape (wpAlpha)}\xspace}
\newcommand{\twpalphaTrue}{{\scshape (wpAlphaTrue)}\xspace}
\newcommand{\twpalphaWedge}{{\scshape (wpAlphaWedge)}\xspace}
\newcommand{\twpalphaVee}{{\scshape (wpAlphaVee)}\xspace}
\newcommand{\twpalphaSigma}{{\scshape (wpAlphaSigma)}\xspace}
\newcommand{\linear}{{\scshape (BetaLinear)}\xspace}
\newcommand{\relu}{{\scshape (BetaRelu)}\xspace}
\newcommand{\sigmoid}{{\scshape (BetaSigmoid)}\xspace}
\newcommand{\tanbeta}{{\scshape (BetaTanh)}\xspace}

\newcommand{\etal}{{\em et al.}\xspace}

\usepackage{tikz}
\usetikzlibrary{positioning}
\usetikzlibrary{shapes.geometric}
\usetikzlibrary{shapes.misc}

\newcommand*\circledbl[1]{\tikz[baseline=(char.base)]{
		\node[shape=circle,fill=blue!20!white,inner sep=1pt] (char) {\textcolor{black}{#1}};}}
\newcommand*\circleO[1]{\tikz[baseline=(char.base)]{
		\node[shape=circle,fill=cyan!15!white,inner sep=1pt] (char) {\textcolor{black}{#1}};}}

\usepackage{listings}
\usepackage{xcolor}

\definecolor{codegreen}{rgb}{0,0.6,0}
\definecolor{codegray}{rgb}{0.5,0.5,0.5}
\definecolor{codepurple}{rgb}{0.58,0,0.82}
\definecolor{backcolour}{rgb}{0.97,0.97,0.98}

\lstdefinestyle{mystyle}{
    backgroundcolor=\color{backcolour},   
    commentstyle=\color{codegreen},
    keywordstyle=\color{magenta},
    numberstyle=\tiny\color{codegray},
    stringstyle=\color{codepurple},
    basicstyle=\ttfamily\footnotesize,
    breakatwhitespace=false,         
    breaklines=true,                 
    captionpos=b,                    
    keepspaces=true,                 
    numbers=left,                    
    numbersep=5pt,                  
    showspaces=false,                
    showstringspaces=false,
    showtabs=false,                  
    tabsize=2
}

\usepackage{subfig}

\lstdefinelanguage{Lambda}{%
  morekeywords={%
    if,then,else,fix 
  },%
  morekeywords={[2]int},   
  otherkeywords={:}, 
  literate={
    {->}{{$\to$}}{2}
    {lambda}{{$\lambda$}}{1}
  },
  basicstyle=\fontsize{9}{9}\ttfamily,
  keywordstyle={[2]},
  	breakatwhitespace=true,         
	keepspaces=true,                 
 	showspaces=false,                
 	showstringspaces=false,
  keepspaces,
  mathescape 
}[keywords,comments,strings]%

\usepackage{mathpartir}

\newcommand{\ccsynth}{\textit{CCSynth}\xspace}

\copyrightyear{2024} 
\acmYear{2024} 
\setcopyright{rightsretained} 
\acmConference[ICSE '24]{2024 IEEE/ACM 46th International Conference on Software Engineering}{April 14--20, 2024}{Lisbon, Portugal}
\acmBooktitle{2024 IEEE/ACM 46th International Conference on Software Engineering (ICSE '24), April 14--20, 2024, Lisbon, Portugal}\acmDOI{10.1145/3597503.3623333}
\acmISBN{979-8-4007-0217-4/24/04}
\begin{document}
\title{Inferring Data Preconditions from Deep Learning Models for Trustworthy Prediction in Deployment}

\author{Shibbir Ahmed}
\affiliation{
	\institution{Dept. of Computer Science\\Iowa State University}
	\city{Ames}
	\state{IA}
	\country{USA}}
\email{shibbir@iastate.edu}

\author{Hongyang Gao}
\affiliation{
	\institution{Dept. of Computer Science\\Iowa State University}
	\city{Ames}
	\state{IA}
	\country{USA}}
\email{hygao@iastate.edu}

\author{Hridesh Rajan}
\affiliation{
	\institution{Dept. of Computer Science\\Iowa State University}
	\city{Ames}
	\state{IA}
	\country{USA}}
\email{hridesh@iastate.edu}

\begin{abstract}
Deep learning models are trained with certain 
assumptions about the data during the development stage 
and then used for prediction in the deployment stage. 
It is important to reason about the trustworthiness of the
model's predictions with unseen data during deployment. 
Existing methods for specifying and verifying traditional software 
are insufficient for this task, as they cannot handle the complexity
of DNN model architecture and expected outcomes.
In this work, we propose a novel technique that uses rules 
derived from neural network computations to infer data preconditions for a DNN model to determine the trustworthiness of its predictions.
Our approach, \deepinfer involves introducing a novel abstraction for a trained DNN model 
that enables weakest precondition reasoning using Dijkstra’s Predicate
Transformer Semantics. 
By deriving rules over the inductive type of neural network 
abstract representation, we can overcome the matrix 
dimensionality issues that arise from the backward non-linear 
computation from the output layer to the input layer. 
We utilize the weakest precondition computation using rules of each kind of activation function to compute layer-wise precondition from the given postcondition on the final output of a deep neural network. 
We extensively evaluated \deepinfer on 29 real-world DNN models using four different datasets collected from five different sources and demonstrated the utility, effectiveness, and performance improvement over closely related work.  
\deepinfer efficiently detects correct and incorrect predictions of high-accuracy models with high recall (0.98) and high F-1 score (0.84) and has significantly improved over the prior technique, \selfchecker. The average runtime overhead of \deepinfer is low, 0.22 sec for all the unseen datasets. We also compared runtime overhead using the same hardware settings and found that \deepinfer is 3.27 times faster than \selfchecker, the state-of-the-art in this area.

\end{abstract}


\maketitle

\section{Introduction}
\label{sec:introduction}

Deep neural networks (DNN) are widely utilized nowadays, including in safety-critical systems. 
A DNN is trained on some data (training data), tested on possibly separate
data (test data), and deployed in production, where they predict output for 
unseen data.
A major challenge is: can we trust the output of a trained DNN on unseen data?
Prior work has referred to these circumstances as data corruption bugs~\cite{Islam2019FSE, Islam2019ICSE}
or conformal constraint violation~\cite{fariha2021conformance,dataPrisomSIGMOD22}.

Prior research on the specification and verification of DNNs has focused on creating abstract representations for the verification of properties such as robustness and fairness~\cite{abstract, abstractdomain, ai2, wang2022interval, aws, huang2017safety, katz2017reluplex, katz2019marabou, urban2020perfectly, biswas2022fairify, mazzucato2021reduced}. However, these works have not addressed the questions of the trustworthiness of DNN outputs~\cite{selfchecker} on unseen data.
Recent studies~\cite{fariha2021conformance, dataPrisomSIGMOD22} have explored techniques for discovering constraints, but they do not consider the DNN's structure in determining these constraints. 
In particular, the conformance constraints approach~\cite{fariha2021conformance} uses the training dataset to establish a "safety envelope" that characterizes the inputs for which the model is expected to make trustworthy predictions. However, this work does not examine whether those conformation constraint violations of the safety envelope can determine correct or incorrect predictions with unseen data in the deployment stage. Our work fills this research gap. 
While many classifiers generate a confidence measure in addition to their class predictions, these measures are often unreliable due to inappropriate calibration~\cite{jiang2018trust, luo2021learning} and may not be sufficient to indicate trust in the classifier's prediction. 
In particular, the application of an activation function to raw numeric prediction values can lead to confidence measures that are not well-calibrated, making them difficult to determine whether the prediction with unseen data during deployment is correct or incorrect. 
\begin{figure*}[h]
	\includegraphics[width=\linewidth,trim={0cm 0cm 0cm 0cm},clip]{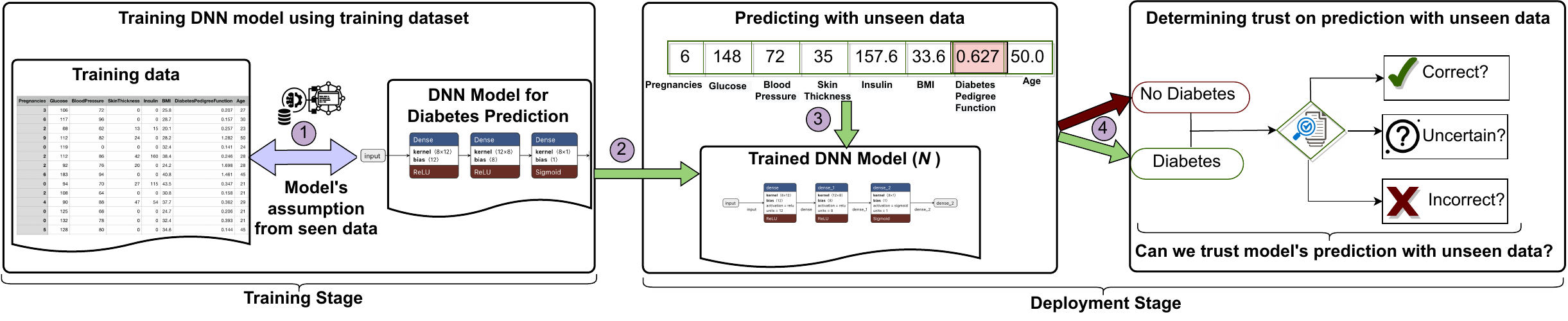}
	\footnotesize
	\caption{An example motivating how we can trust a model's prediction with unseen data in the deployment stage} 
	\label{fig:motivExample}
\end{figure*}

Recently, Xiao \etal proposed a technique, \selfchecker~\cite{selfchecker} that computes the similarity between layer features of test instances and the samples in the training set, using kernel density estimation (KDE) to detect misclassifications by the model in deployment. This technique has limitations, such as being restricted to the capability of computing density function from specific training and test data and the selected combination of layers with certain activation functions. Therefore, \selfchecker incurs a significant runtime overhead to compute KDEs for different combinations of layers for each class in training and deployment modules for all
training and test datasets.
We address these shortcomings of the state-of-the-art techniques and aim to develop a technique that infers DNN model's assumption on training data and utilizes that inferred assumption during the deployment stage to determine correct or incorrect prediction, therefore implying trust in prediction with unseen data.

In this work, we provide a novel approach \deepinfer for reasoning about a DNN's 
prediction with unseen data by inferring data preconditions from the 
DNN model, i.e., structure of the DNN and trained parameters.
The technical contributions of our approach include:
a novel abstraction of DNN, including conditions, 
a weakest precondition ({\em wp}) calculus~\cite{hoare1987weakest} for DNNs, 
and an algorithm that utilizes derived rules from the DNN abstraction and layer-wise computations to infer data preconditions 
and determine the model's correct or incorrect prediction.
Starting with the conditions that should hold on the output of the DNN ({\em postconditions}),
our {\em wp} rules provide mechanisms to compute conditions on the input of that layer ({\em preconditions}).
Since the output of one layer ($N$) is fed to the input of the next layer ($N+1$) in a DNN, 
our approach then uses the preconditions of the $N+1$ layer as postconditions of 
the previous layer $N$. 
The precondition of the first layer, also called the input layer, in the DNN 
are {\em data preconditions}.
The challenge in formulating {\em wp} rules lies in handling multiple layers with 
hidden non-linearities due to the architecture of the DNNs.

To evaluate our approach, we utilize 29 real-world models and 4 different datasets collected from prior research~\cite{udeshi,zhang,aggarwal,biswas2022fairify} and Kaggle~\cite{kaggle} to 
answer three research questions. 
We investigated whether data precondition violations determine incorrect model prediction. 
We also measure how effective \deepinfer is to imply trustworthiness in the model’s prediction and compare against closely related work using their evaluation metrics~\cite{selfchecker}. We determine the performance, especially the runtime overhead of \deepinfer and compared it with the state-of-the-art using unseen data during deployment.
Our key results are:
 \textbf{\deepinfer implies that data precondition violations and incorrect model prediction 
 are highly correlated ($pcc = 0.88$)}, where $pcc$ denotes Pearson correlation coefficient. Also, \textbf{the data precondition satisfaction and correct model prediction are strongly correlated ($pcc = 0.98$)}. 
 \textbf{\deepinfer effectively implies the correct and incorrect prediction of higher accuracy models with recall (0.98) and F-1 score (0.84), compared to prior work \selfchecker with recall (0.59) and F-1 score (0.52)}. The average runtime overhead of \deepinfer is fairly minimal (0.22 sec for the entire test data). \textbf{Our proposed approach, \deepinfer is 3.27 times faster during deployment than \selfchecker, state-of-the-art in this area.} 

In summary, this work makes the following contributions:
\begin{itemize}

    \item a novel abstraction for trained DNN that incorporates pre and postconditions 
        as predicate vectors for each layer; 
    \item a weakest precondition calculus for the DNN abstraction 
    that overcomes challenges due to non-linearities introduced by the DNN architecture; 
    \item a novel technique for computing data preconditions from DNN models after training and utilizing those inferred preconditions for implying trust in the model’s prediction during the deployment stage ; 
    \item a detailed evaluation with publicly available datasets and models to demonstrate the utility, efficiency, and performance of \deepinfer with an open-source implementation~\cite{DIrepo} that can be leveraged by future research in explainable software engineering for machine learning.
\end{itemize}

\section{Motivation}
\label{sec:background}
We are aware that a DNN model's prediction could be correct or incorrect, but it is important to know how trustworthy the model's prediction is for unseen data during the deployment stage.
To motivate our objectives, let us consider a deep neural network model in Fig.~\ref{fig:motivExample}. The first layer, i.e., the DNN model's input layer, receives the input from training data, compiles it and produces the output (\circledbl{1}). Then, the next layers receive the output from the previous ones as input. The model compiles the input data, evaluates it, predicts the output, and delivers it to the deployment stage (\circledbl{2}). 
This model has been trained for the PIMA diabetes dataset with eight features 
for whether a patient has diabetes. Although the model's accuracy is ~77\%, when we get the output from the model, we do not really know how confident the model is for that output. In some cases, the model could be confidently incorrect. So, this model's prediction with an unseen data during the deployment stage might be correct or incorrect. For instance, during the deployment stage, unseen data is fed to the trained DNN model (\circledbl{3}), which predicts whether the patient with that particular data point has diabetes or no diabetes (\circledbl{4}). It is necessary to determine whether the model's prediction is correct and to trust this prediction or its prediction is incorrect and not to trust it with such unseen data points during the deployment stage.
\begin{figure*}[h]
	\includegraphics[width=\linewidth,trim={0cm 0cm 0cm 0cm},clip]{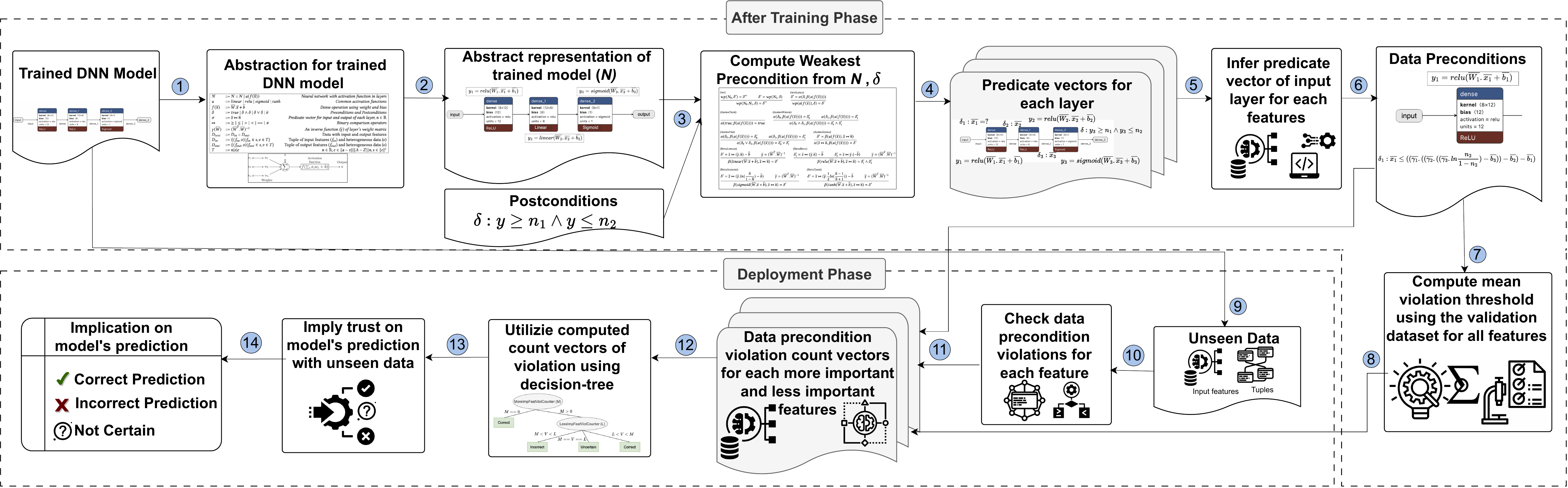}
	\footnotesize
	\caption{Overview diagram depicting the technique of data precondition inference from a trained DNN model after the training phase and how those are utilized in the deployment stage for implying trust in the model's prediction using unseen data} 
	\label{fig:overview}
\end{figure*}
The growing prevalence of Deep Neural Networks (DNNs) in critical domains highlights the importance of ensuring the trustworthiness of their outputs. Despite their high accuracy, DNNs are still prone to prediction errors, and in applications such as autonomous vehicles 
and medical diagnosis, etc. It is reported that Uber's fatal self-driving crash was caused by software detecting objects on the road~\cite{Uberaccident} and AI models for health care that predict disease are not as accurate as suggested in reports~\cite{medicine}. 
Therefore, making the DNN black-box model explainable and determining correct, incorrect, or uncertain predictions during deployment is crucial. 

\textbf{Problem formulation:}
Given a trained DNN model and an unseen data instance, our goal is to  
derive preconditions from a trained DNN model's assumptions about the training data after the training stage, and leverage inferred data preconditions from the model to precisely determine whether a  prediction by the DNN model with unseen data during deployment is correct or incorrect, or uncertain. By addressing the challenges posed by non-linear computation functions in DNN model and the variability of weights, biases, inputs, and outputs, our work aims to provide an efficient solution and significantly improved technique over state-of-art for ensuring trust in the DNN model's prediction in real-world applications.

\section{\deepinfer Approach}
\label{sec:approach}
We present an overview diagram in~\figref{fig:overview} illustrating our proposed technique \deepinfer. The top portion of the diagram depicts how data preconditions are inferred from a trained DNN model after the training phase. In the bottom portion, we depict how the inferred data preconditions are utilized for determining the trustworthiness of the model's prediction using unseen data during deployment. At first, we utilize a trained DNN model for the novel abstraction with layers and activation functions incorporating preconditions and postconditions (\circleO{1}). Then, we represent a neural network with activation function operations inside layers (\circleO{2}). We compute the weakest preconditions from the abstract representation of the trained model ($N$) and postcondition ($\delta$) (\circleO{3}). Then, we determine the predicate vectors for each layer utilizing the computed weakest preconditions from layer-wise operations (\circleO{4}). From (\circleO{5}), we infer the input layer's predicate vector for each feature. Therefore, we obtain the data preconditions using the trained model once after the training phase (\circleO{6}). Then, we compute mean data precondition violations for all features using the entire validation dataset, which
serve as a threshold
(\circleO{7}).
In the deployment phase, \deepinfer utilizes the trained DNN model and obtained data preconditions for determining trust in the model's prediction with an unseen data point (\circleO{9}). Next, we check the data precondition violations (\circleO{10}) for each feature using the violation threshold (\circleO{8}) for that data point (\circleO{11}).  
Furthermore, we utilize the computed count vectors of the violation using a decision-tree-based approach (\circleO{12}). To that extent, we determine the trustworthiness of the model's prediction with unseen data (\circleO{13}). Finally, \deepinfer determines whether the model's prediction is correct and we can trust it or incorrect and not certain, and we can not rely on that prediction with unseen data during the deployment stage (\circleO{14}).

\subsection{Abstract representation of a DNN model}
\label{abstract}
We propose a novel abstraction for trained DNNs that incorporates pre and postconditions as predicate vectors for each layer. Let us consider the following grammar for representing DNN depicted in~\figref{fig:wpgrammar}. 
\begin{figure}[htbp]
\resizebox{\columnwidth}{!}{
\begin{tabular}{ |l l r|}
\hline
 $N$ & ::= $N::N$ | $a(f(\overline{x}))$ & \textit{Neural network with activation function} \\
 $a$ & ::= $linear$ | $relu$ | $sigmoid$ | $tanh$  & \textit{Common activation functions} \\
 $f(\overline{x})$ & ::= $\overline{W}.\overline{x}+\overline{b}$ &  \textit{Dense operation using weight and bias}   \\
 $\delta$ & ::= $true$ | $\delta \wedge \delta$  | $\delta \vee \delta$ | $\overline{\sigma}$  & \textit{Preconditions and Postconditions} \\ 
 $\sigma$ & ::= $\overline{z} \bowtie \overline{n}$  & \textit{Predicate vector of layer's input/output ($\overline{z}$), $n \in \mathbb{R}$} 
 \\
 
 $\bowtie$ & ::= $\ge | \le | > | < | == | \ne$  & \textit{Binary comparison operators} \\
 $\gamma(\overline{W})$ & ::= $(\overline{W}^T.\overline{W})^{-1}$   & \textit{Inverse function ($\overline{\gamma}$) of layer's weight matrix} \\
$D_{test}$ & ::= $D_{in} :: D_{out} $ &  Data with input and output features \\  
$D_{in}$ & ::= $\{ (f_{in},v) | f_{in} \in s , v \in T\}$ & Tuple of input features ($f_{in}$) and data ($v$) \\
$D_{out}$ & ::= $\{ (f_{out},v) | f_{out} \in s , v \in T\}$ & Tuple of output features ($f_{out}$) and data ($v$) \\
  $T$ & ::= $n | s | c$ & $n \in \mathbb{R}, c \in [a-z]|[A-Z]|n, s \in [c]^*$  \\
 \hline
\end{tabular}
}
\caption{Grammar representing Neural network, preconditions, and postconditions}
\label{fig:wpgrammar}
\end{figure}

Let us consider the \texttt{Dense} layer computation denoting $f(\overline{x})$. In the grammar, 
we denote $N$ as a neural network with activation function $a(f(\overline{x}))$ in layers. In this computation, the function is based on the neuron’s weights, and bias where one weight is assigned to each component of the input $(\overline{x})$ with corresponding weight $(\overline{W})$ and bias $(\overline{b})$ in each layer. 
We consider some common activation functions~\cite{wang2022interval} used in deep learning programs such as \texttt{linear, ReLU, sigmoid, tanh}. We consider each layer's output and input vector as $\overline{z}$ and the predicate as $\bowtie \overline{n}$, where $n \in R$ and $\bowtie$ represent the logical comparison operators. Here, $\overline{\gamma}$ denotes an inverse function of a layer’s weight matrix nonlinear computation. We represent the test dataset ($D_{test}$) as a tuple of features and data. 

\subsection{Computing weakest preconditions from abstract representation of a DNN model}
\label{sec:wp}
\begin{figure*}
\small
\setlength{\parskip}{.1cm}
\setlength{\belowcaptionskip}{.05cm}
\[
\inferrule[(wp)] 
{\wpre (N_{0}, \delta') = \delta'' \;\; \delta' = \wpre (N_{1}, \delta)} 
{\wpre (N_{0}.N_{1},\delta) 
= \delta''}   
\;\;\;\;\;\;\
\inferrule[(wpAlpha)] 
{ \delta' = \alpha (\delta, \beta (a(f(\overline{x}))))} 
{\wpre (a(f(\overline{x})), \delta) = \delta'}  
\;\;\;\;\;\;\
\inferrule[(AlphaTrue)] 
{\;\;\;\;\;} 
{\alpha (true, \beta (a(f(\overline{x})))) = true}  
\;\;\;\;\;\;\
\inferrule[(AlphaWedge)] 
{\alpha (\delta_{0}, \beta (a(f(\overline{x})))) = \delta_0' \;\; \alpha (\delta_{1}, \beta (a(f(\overline{x})))) = \delta_1'}  
{\alpha (\delta_0 \wedge \delta_1, \beta (a(f(\overline{x})))) = \delta_0' \wedge \delta_1'}  
\]

\[
\inferrule[(AlphaVee)] 
{\alpha (\delta_{0}, \beta (a(f(\overline{x})))) = \delta_0' \;\;\;\; \alpha (\delta_{1}, \beta (a(f(\overline{x})))) = \delta_1'}
{\alpha (\delta_0 \vee \delta_1, \beta (a(f(\overline{x})))) = \delta_0' \vee \delta_1'}  
\;\;\;\
\inferrule[(AlphaSigma)] 
{\delta' =\beta (a(f(\overline{x})), \overline{z} \bowtie \overline{n}) } 
{\alpha (\overline{z} \bowtie \overline{n}, \beta (a(f(\overline{x})))) = \delta '}
\;\;\;\;\;\;\
\inferrule[(BetaRelu)] 
{ \delta_1' =\overline{z} \bowtie (\overline{\gamma}.\overline{n}) - \overline{b} \;\;\;\;\; \delta_2' =\overline{z} \bowtie \overline{\gamma}.(- \overline{b}) \;\;\;\; \overline{\gamma} = (\overline{W}^T.\overline{W})^{-1}.(\overline{W^T})} 
{\beta ( relu(\overline{W}.\overline{x}+\overline{b}), \overline{z} \bowtie \overline{n}) = \delta_1' \wedge \delta_2'}  
\]
\[
\inferrule[(BetaLinear)] 
{ \delta' =\overline{z} \bowtie (\overline{\gamma}.\overline{n}) - \overline{b}  \;\; \overline{\gamma} = (\overline{W}^T.\overline{W})^{-1}.(\overline{W^T})}
{\beta ( linear(\overline{W}.\overline{x}+\overline{b}), \overline{z} \bowtie \overline{n}) = \delta'}  
\;
\inferrule[(BetaSigmoid)] 
{\delta' =\overline{z} \bowtie (\overline{\gamma}.ln (\frac{\overline{n}}{1-\overline{n}}) - \overline{b}) \;\; \overline{\gamma} = (\overline{W}^T.\overline{W})^{-1}.(\overline{W^T})} 
{\beta ( sigmoid(\overline{W}.\overline{x}+\overline{b}), \overline{z} \bowtie \overline{n}) = \delta' } 
\;\;
\inferrule[(BetaTanh)] 
{ \delta' =\overline{z} \bowtie (\overline{\gamma}.\frac{1}{2}ln (\frac{\overline{n}-1}{\overline{n}+1}) - \overline{b}) \;\; \overline{\gamma} = (\overline{W}^T.\overline{W})^{-1}.(\overline{W^T})} 
{\beta ( tanh(\overline{W}.\overline{x}+\overline{b}), \overline{z} \bowtie \overline{n}) = \delta'}  
\]
\caption{Rules for computing $\wpre$ over inductive type $N$, $\alpha$ over inductive type $\delta$, $\beta$ over inductive type $a(f(\overline{x}))$ 
}
\label{tab:wptype}
\end{figure*}

To compute the weakest preconditions from the abstract representation of a DNN model ($N$), we consider a 
postcondition $\delta$ as the expected DNN model's output.  
As edges from one layer connect the neural network to another layer, we denote $\overline{z}$ as the input/output vector for the intermediate layers. 
We consider $\overline{y}$ the output of the last layer, and $n$ is the prediction interval for a DNN model’s prediction. Here, $\bar{x}$ is the input vector to the first layer. It  is considered as data precondition.
We follow Dijkstra's predicate transformer semantics rules in the form of $\wpre(N, \delta)$, where $N$ is a program statement involving a DNN layer computation using activation functions as represented in the grammar, and $\delta$ is a postcondition on the program state. This transformer rule defines the weakest predicate, which holds the model statement before executing $N$ to guarantee that the postcondition $\delta$ holds after $N$ terminates.
\deepinfer computes data precondition for a model using the defined rules in~\figref{tab:wptype}. We compute the data precondition, which is obtained recursively by following these rules from the last layer until the first layer of a DNN model.
Therefore, the computation of the data precondition from a DNN model is done recursively for a given representation $N$ from the DNN model and postcondition $\delta$ using the rules illustrated in~\figref{tab:wptype}.
Here, the rules \twp, \twpalpha represent recursion over inductive type $N$ by the function \wpre, eventually satisfying base cases of $N$.
These base cases of \wpre use $\alpha$ to compute the precondition where $\alpha$ does recursion over the cases of the inductive type $\delta$ represented using rules \twpalphaTrue, \twpalphaWedge , \twpalphaVee, \twpalphaSigma illustrated in~\figref{tab:wptype}. 
Again, base cases of $\alpha$ use $\beta$ to compute the \wpre for the cases of the activation function ($a$). 
For instance, for \texttt{ReLU} activation function, we compute $\beta$ using the computation with weight and bias of a layer as follows,
\[
relu(f(\overline{x})) = f(relu(\overline{W}.\overline{x}+\overline{b})) =
\begin{cases} 
0, & (\overline{W}.\overline{x}+\overline{b}) < 0\\
\overline{W}.\overline{x}+\overline{b}, & (\overline{W}.\overline{x}+\overline{b}) \ge 0
\end{cases}
\]
We solve this non-linear equation of $relu(\overline{W}.\overline{x}+\overline{b})$ for postcondition ($\overline{z} \bowtie \overline{n}$)
and obtain the precondition 
as stated in the \relu rule in~\figref{tab:wptype}. Similary, for other kinds of activation functions, we have derived $\beta$ rules e.g., \linear, \sigmoid, \tanbeta rules illustrated in~\figref{tab:wptype}. 
The derivation details of each kind of those rules are in the appendix of open-source repository~\cite{DIrepo}.
\[
\beta ( relu(W.\overline{x}+b), \overline{z} \bowtie \overline{n}) \equiv \overline{z} \bowtie ((\overline{\gamma}.\overline{n}) - \overline{b}) \wedge \overline{z} \bowtie \overline{\gamma}.(- \overline{b}) , \overline{\gamma} = (\overline{W}^T.\overline{W})^{-1}
\]

Next, we describe the challenges towards layer-wise weakest precondition reasoning using a DNN model.

\subsection{Layer-wise weakest precondition reasoning}
\label{sec:layerwisewp}
In order to obtain \wpre by asserting the model statement using postcondition from layer to layer, there are some challenges. First, the layer function computation using the activation function is not always linear. Different non-linear activation functions operate using weight and bias along with the input in each layer computation. For instance, \texttt{sigmoid} activation function computes ($\sigma(x) = \frac{1}{1+e^{-x}}$), \texttt{tanh} activation function computes ($tanh(x) = \frac{2}{1+e^{-2x}}$ -1),  
\texttt{ReLU} activation function computes ($relu(x) = x , x \ge 0 | 0, x <0$), \texttt{ELU} activation function computes ($elu(x) = x , x \ge 0 | e^x -1, x <0$), 
\texttt{softmax} activation function computes ($softmax(x_i) = \frac{e^{x_i}}{\sum_{j=1}^{n} e^{x_j}}$)~\cite{wang2022interval}, etc.
\begin{figure}[!h]
	\includegraphics[width=\columnwidth,trim={1.5cm 1.9cm 0cm 1cm},clip]{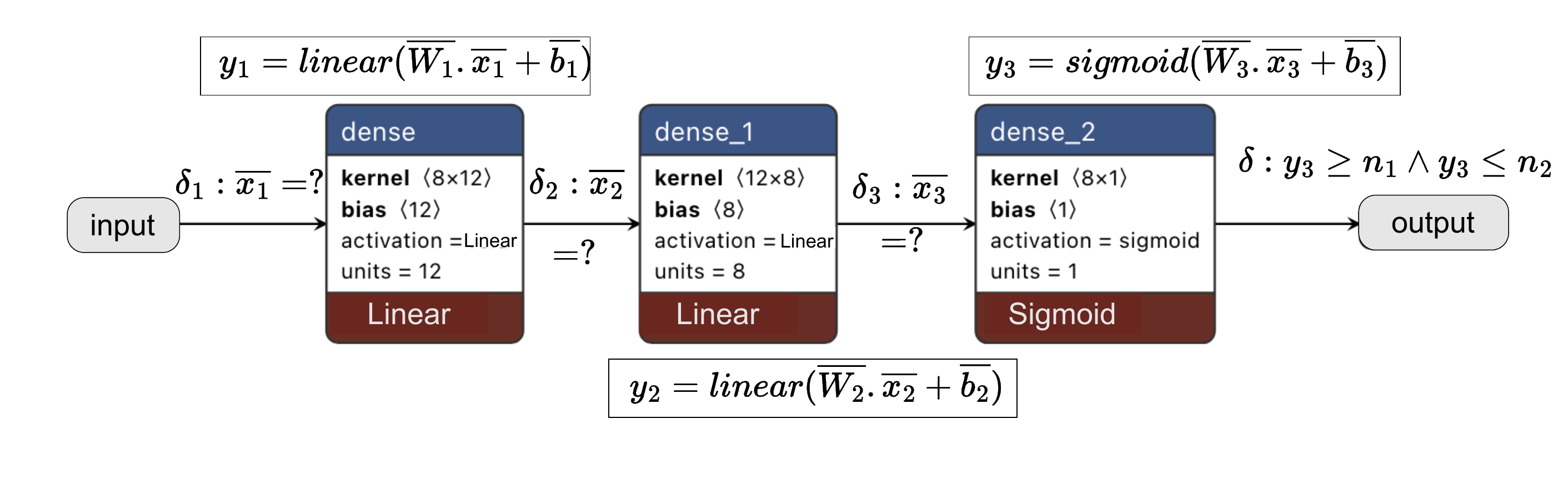}
	\footnotesize
	\caption{Data precondition ($\delta_1$) computation from an example DNN model ($N$) with 3 layers and postcondition ($\delta$) } 
	\label{fig:wpModel3}
\end{figure}

Second, there is a challenge to tackle the variability of the matrix dimension of weight, bias, input, and output in each layer. 
For instance, an example model (in Fig.~\ref{fig:wpModel3})
contains 3 \dense layers which perform \texttt{linear, linear, sigmoid} activation function computation using weight and bias vector with input in each layer. To obtain the layer-wise \wpre, the dimension of weight and bias matrices should be taken into account.
The dimension of weight matrices varies from layer to layer in the network. 
As the weight vector ($\overline{W}$) is multiplied by the input vector ($\overline{X}$), the dimension must be consistent with the bias vector ($\overline{b}$) and output ($\overline{y}$) in forward propagation. 
In terms of backward computation, it is challenging to get the appropriate matrix dimension on the precondition of input data in each layer. In~\figref{fig:wpModel3}, the dimension of weight, bias, input, and output of last layer is ($1 \times 8$), ($1 \times 1$), ($8 \times 1$), ($1 \times 1$) respectively. In the second layer, the dimension of weight, bias, input, and output is ($8 \times 12$), ($8 \times 1$), ($12 \times 1$), ($8\times 1$), respectively. In the first layer, the dimension of weight, bias, input, and output is ($12 \times 8$), ($12 \times 1$), ($8 \times 1$), ($12\times 1$), respectively. If we assert using postcondition with a single dimension of $\delta$, as a data precondition in the first layer should be a dimension of $8 \times 1$ in this scenario. We encounter here that the weight, bias, input, and output of each layer appear non-linearly in the equations of the activation function, where there are nonlinear constraints among the parameters. To address these challenges, we have adopted the least square solution~\cite{jefferys1980method} for nonlinear activation computation. One of our contributions is to derive \wpre rules for each kind of activation function (shown in Fig.~\ref{tab:wptype}) for layer-wise weakest precondition reasoning. 

Next, we describe the weakest precondition computation of a DNN model to infer data preconditions of the input layer using the derived rules (in~\figref{tab:wptype}). Our approach is generalized to a DNN with any number of hidden layers with linear or non-linear activation functions. For simplicity, we demonstrate the \wpre computation process using derived rules with a canonical example DNN model.

\subsection{Infer data preconditions of the input layer} 
\label{sec:datapre}
For computing the weakest precondition using DNN models, we take the statements from the model structure consisting of layers with input dimensions, number of output, activation function, etc. We consider the prediction interval ($n$) as the postcondition. The rationale behind choosing prediction interval as a postcondition to DNN classification or regression model is that it~\cite {kuhn2013applied} provides how good the model prediction is.
Also, the prediction interval helps gauge the weight of evidence available when comparing models. Prediction intervals facilitate trade-offs between models favoring less complex or more interpretable models~\cite {breiman2001statistical}. 

To infer data preconditions, starting from the last layer statement, we assert using the \wpre equation to determine the weakest precondition. 
The equation of $Dense$ layer is as follows: 
\[
output = activation(dot(input, weight) + bias)
\]

So, for given postcondition $\delta: y \le n$,  the statement $S_3$ can be written for output of last layer ($y_3$) with corresponding weight ($\overline{w_3}$) and bias ($\overline{b_3}$) as,
\[
y_3 = sigmoid (\overline{w_3^T}.\overline{x_3} + \overline{b_3})
\]

We have \dense layers with \texttt{linear, linear, sigmoid} activation functions for this example. Now, for given neural network ($N$) and postcondition ($\delta$),
\[
N : linear(\overline{W_1}.\overline{x_1}+\overline{b_1}). linear(\overline{W_2}.\overline{x_2}+\overline{b_2}).sigmoid(\overline{W_3}.\overline{x_3}+\overline{b_3}) ;
\]
\[
\delta : y_3 \ge n_1 \wedge y_3 \le n_2
\]

Our proposed technique is generalized to DNN models with multiple layers. For example, a DNN model presented in Fig.~\ref{fig:wpModel3} has 3 layers and different activation functions. In that model, the output layer has a single class, i.e., the output value $y \in \mathbf{R}$. The given postcondition is an instance of ($\delta \wedge \delta$) and will be in the range between $[n_1, n_2]$.  
Now, we utilize \wpre rules over $N$ and $\delta$ using \twp, \twpalpha rules to get the precondition for this multiple layer neural network as follows,
\[
\delta_1 = \wpre (N, \delta) = \wpre (N_0.N_1, \delta) \equiv
\]
\[
\wpre (linear(\overline{W_1}.\overline{x_1}+\overline{b_1}). linear(\overline{W_2}.\overline{x_2}+\overline{b_2}).sigmoid(\overline{W_3}.\overline{x_3}+\overline{b_3}), \delta)
\]
\[
\delta_1 = \wpre (linear(\overline{W_1}.\overline{x_1}+\overline{b_1}, \delta_2) ; 
\]
\[
\delta_2 = \wpre (linear(\overline{W_2}.\overline{x_2}+\overline{b_2}).sigmoid(\overline{W_3}.\overline{x_3}+\overline{b_3}), \delta);
\]
\[
\delta_2 = \wpre (linear(\overline{W_2}.\overline{x_2}+\overline{b_2}), \delta_3);
\delta_3 = \wpre (sigmoid(\overline{W_3}.\overline{x_3}+\overline{b_3}), \delta)
\]

Then, we apply \twpalphaSigma, \twpalphaWedge, \sigmoid rules consecutively to get the precondition as follows,
\[
\delta_3 = \wpre (sigmoid(\overline{W_3}.\overline{x_3}+\overline{b_3}), \delta) 
\equiv \alpha (\delta_3, \beta(sigmoid(\overline{W_3}.\overline{x_3}+\overline{b_3})))
\]
\[
\equiv \beta(sigmoid(\overline{W_3}.\overline{x_3}+\overline{b_3}), \delta_3) 
\equiv \beta(sigmoid(\overline{W_3}.\overline{x_3}+\overline{b_3}), y_3 \ge n_1 \wedge y_3 \le n_2)
\]
\[
\equiv \overline{x_3} \ge ((\overline{\gamma_3}.ln \frac{n_1}{1-n_1}) - \overline{b_3}) \wedge \overline{x_3} \le ((\overline{\gamma_3}.ln \frac{n_2}{1-n_2}) - \overline{b_3})
\]

Here, $\overline{x_3}$ is an array of input that has been obtained from the second layer and fed into the third layer, and the predicate of $\overline{x_3}$ denotes the precondition of the data in layer 3, which is a postcondition of layer 2. Here, $\overline{\gamma_3}$ is an inverse function of the layer's weight matrix ($\overline{W_3}$). 
Then, we obtain $\delta_2$ similarly using the \wpre rules \twpalpha, \twpalphaWedge, \linear consecutively,
\[
\delta_2 = \wpre (linear(\overline{W_2}.\overline{x_2}+\overline{b_2}),\delta_3) 
\]
\[
\equiv \alpha (\delta_3, \beta(linear(\overline{W_2}.\overline{x_2}+\overline{b_2})))
\equiv \beta(linear(\overline{W_2}.\overline{x_2}+\overline{b_2}), \delta_3) 
\]
\[
\equiv \beta(linear(\overline{W_2}.\overline{x_2}+\overline{b_2}), \overline{x_3} \ge ((\overline{\gamma_3}.ln \frac{n_1}{1-n_1}) - \overline{b_3}) \wedge 
\]
\[
\overline{x_3} \le ((\overline{\gamma_3}.ln \frac{n_2}{1-n_2}) - \overline{b_3}))
\equiv \overline{x_2} \ge ((\overline{\gamma_2}.((\overline{\gamma_3}.ln \frac{n_1}{1-n_1}) - \overline{b_3})) - \overline{b_2}) \wedge
\]
\[
\overline{x_2} \le ((\overline{\gamma_2}.((\overline{\gamma_3}.ln \frac{n_2}{1-n_2}) - \overline{b_3})) - \overline{b_2}),
\overline{\gamma_2} = (\overline{W_2^T}.\overline{W_2})^{-1}.(\overline{W_2^T}) 
\]
In this step, we obtain the precondition which is an array of the input ($\overline{x_2}$) that has been obtained from the first layer and fed into the second layer, and the predicate of $\overline{x_2}$ denotes the precondition of the input in layer 2, which is a postcondition of layer 1. After asserting with this postcondition, we obtain $\delta_1$ similarly using the \wpre rules \twpalpha, \twpalphaWedge, \linear,
\[
\delta_1 = \wpre (linear(\overline{W_1}.\overline{x_1}+\overline{b_1}),\delta_2) 
\equiv \alpha (\delta_2, \beta(linear(\overline{W_1}.\overline{x_1}+\overline{b_1})))
\]
\[
\equiv \beta(linear(\overline{W_1}.\overline{x_1}+\overline{b_1}), \delta_2) 
\]
\[
\equiv \overline{x_1} \ge ((\overline{\gamma_1}.((\overline{\gamma_2}.((\overline{\gamma_3}.ln \frac{n_1}{1-n_1}) - \overline{b_3})) - \overline{b_2}) - \overline{b_1})
\wedge
\]
\[
\overline{x_1} \le ((\overline{\gamma_1}.((\overline{\gamma_2}.((\overline{\gamma_3}.ln \frac{n_2}{1-n_2}) - \overline{b_3})) - \overline{b_2}) - \overline{b_1})
\]
Finally, we obtain the precondition, which is an array of the data ($\overline{x_1}$) for each feature that has been assumed by this DNN with multiple layers where $\overline{\gamma_1} = (\overline{W_1^T}.\overline{W_1})^{-1}.(\overline{W_1^T})$.
In our proposed technique, the entire process of data precondition inference from a DNN model is automated and generalized for other models which is performed after the training stage. Next, we discuss how we utilize inferred data preconditions for determining the trustworthiness of the model’s prediction using unseen data. 
\subsection{Implying trustworthiness on the model's prediction using inferred data preconditions}
\label{sec:implytrust}
Regarding the design choice, we determine the data preconditions for the inputs to the first layer in a DNN model. These data preconditions for the inputs to a DNN model indicate the trained model's assumption about the data. Furthermore, these input data preconditions must hold true for the data before it is fed to the model, which is important for its prediction. 
Prior work regarding the conformance constraints approach~\cite{fariha2021conformance} uses the training dataset to establish a "safety envelope" that characterizes the inputs and demonstrates that conformal constraint violation is related to a model's trustworthy predictions.
We leverage a similar notion in our approach that the violation of obtained data preconditions for the input to a DNN model indicates the trustworthiness of the model's prediction.

The overall process has two parts shown in Algorithm~\ref{algoUnseen2}. 
The procedure \textsc{computeThreshold} computes the 
violation threshold for input features using the validation set, 
and \textsc{checkPrediction} uses these computed values to check
prediction for unseen data.
Given the neural network representation $N$ and the postcondition $\delta$,
the first step is to acquire the data preconditions (line 2),
set of input features, and
data points from the validation dataset $D_{\text{test}}$ (lines 3-5).
The algorithm proceeds by collecting feature-wise violations using 
the helper procedure on lines 11--18, which checks precondition violation
for each input in the validation set and accumulates the precondition 
violations by features.
Finally, we calculate the mean number of data precondition violations for 
all features ($V$), which serve as a threshold (on line 9).
\begin{figure}[!h]
\centering
	\includegraphics[width=3.2in,trim={0cm 0cm 0cm 0cm},clip]{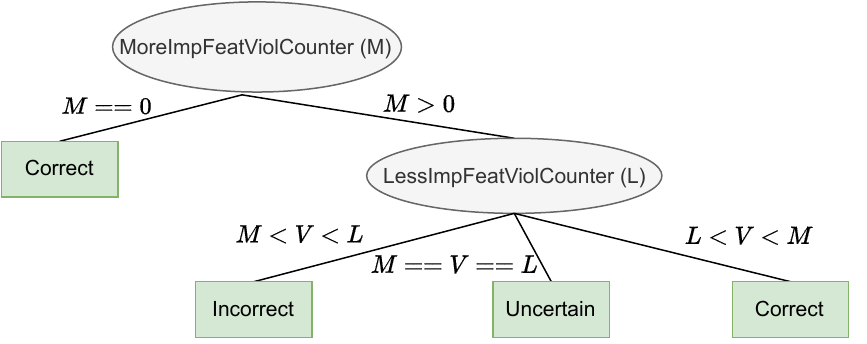}
	\footnotesize
	\caption{Utilizing computed count vectors of the data precondition violations using decision-tree} 
	\label{fig:decisionTree}
\end{figure}
For the unseen data, procedure \textsc{checkPrediction} computes 
the violation count for each feature (line 21). 
Next, for each feature the procedure checks whether the number of violations
are above  ($L$) or below ($M$) the violation threshold.
To be more specific regarding the design choice of the decision tree (in \figref{fig:decisionTree}) of data preconditions violation, we have utilized more feature violations and fewer feature violations as indicative of the model's correct and incorrect prediction.
The decision tree logic is in \figref{fig:decisionTree}. 
First leaf (from the left) of this decision tree is immediate, 
if there are no more violations compared to the threshold 
then the model's prediction is correct.
If $L==V==M$, then the procedure is unsure about the output of the model
and therefore we assign it uncertain (leaf 3).
If $L<V<M$, then there are more features for which the precondition violation 
is below the threshold and fewer features for which the violation is above. 
That means the overall violation is less, leading to correct prediction (leaf 4).
Finally, if $M<V<L$, there are more precondition 
violations above the threshold, and thus the model output 
is incorrect (leaf 2). To be more specific regarding the design choice of the decision-tree of data preconditions violation is that we utilized the more feature violations and less feature violations as indicative of the model's correct and incorrect prediction.

\begin{algorithm}[h!t]
\footnotesize
	\caption{Data Precondition Violation Procedure 
	}\label{algoUnseen2}
	\begin{algorithmic}[1]
		\Procedure{computeThreshold($N$, $D_{test}, \delta$)}{}
        \State $\delta' \gets \wpre (N,\delta)$  
 \algorithmiccomment{Obtain data precondition given $N$ and postcondition $\delta$}
      \State $\overline{f} \gets \mathbf{dom} (D_{in})$ 
      \algorithmiccomment{Set of input features, $D_{in}  \in D_{test}$}
       \State $d \gets \mathbf{range} (D_{in})$ \algorithmiccomment{Set of data points, $v \in D_{in}$}
     \State $wp_{vDict} \gets \varnothing$ 
    \State $\overline{v} \gets \Call{collectFeatureWiseViolations}{d, \overline{f}, \delta'}$   
    \ForEach {$i$ $\in$ $|\overline{f}|$} 
     \State $wp_{vDict} \gets wp_{vDict} \cup \{\langle i, \overline{v}[i] \rangle \}$
     \EndFor
\State $V \gets \frac{1}{n}\sum_{n=1}^{n=|\overline{f}|}(v_{wp}) | v_{wp} \in wp_{vDict}$ \algorithmiccomment{Mean violation threshold}
\State \Return $V, \delta', \overline{f}$ 
\EndProcedure
\Procedure{collectFeatureWiseViolations($d$, $\overline{f}$, $\delta'$)}{}
     \State $\overline{v} \gets 0$    \algorithmiccomment{Violation count array indexed by features}
     \ForEach {$\overline{t}$ $\epsilon$ $d$}
        \State $\overline{v_{wp}} \gets \delta'( \overline{t} )$ \algorithmiccomment{Check precondition violation for input}
        \ForEach {$i$ $\in$ $|\overline{f}|$}  \algorithmiccomment{Collect precondition violation for each feature}
            \If {$\overline{v_{wp}}[i]$}
               \State $\overline{v}(i) \gets \overline{v}(i)+1$
               \EndIf
        \EndFor
     \EndFor
    \State \Return $\overline{v}$ 
\EndProcedure  
\Procedure{checkPrediction ($d$, $V$, $\delta', \overline{f}$)}{}
    \State $M, L \gets  0$, $wp_{warn} \gets \varnothing$
    \State $\overline{v} \gets \Call{collectFeatureWiseViolations}{d, \overline{f}, \delta'}$ 
    \ForEach {$i$ $\in$ $|\overline{f}|$} 
         \If {($\overline{v}[i] \le V$)} \algorithmiccomment{Compare violation count with threshold}
             \State $M \gets M+1$
         \Else
             \State $L \gets L+1$
         \EndIf
    \EndFor
    \State $wp_{warn} \gets decisionTree(M,V,L) $ \algorithmiccomment{Correct/incorrect/uncertain?}
    \State \Return $wp_{warn}$ 
\EndProcedure
\end{algorithmic}
\end{algorithm}

{\em Time Complexity.\ }
The procedure \textsc{checkPrediction} doesn't compute over the DNN. 
It uses preconditions computed by the procedure \textsc{computeThreshold} that 
{\em runs once per DNN} after training.
The time complexity of the procedure \textsc{computeThreshold} is dominated by the \wpre function,
whose complexity is akin to the back-propagation algorithm of a FCNN. 
The time complexity is primarily determined by matrix multiplications,
that has the complexity $O(n^{log_2 7})$ for Strassen's method~\cite{cenk2017arithmetic}. 
The time complexity of \wpre is $O(|N| + n^{log_2 7})$ 
where, $|N|$ is the length of layers of model and $n$ is the dimension of the weight matrix.
The time complexity of \textsc{checkPrediction} is $O(|d|.|\overline{f}|)$ where $|d|$ is the size of unseen data and $|\overline{f}|$ is the number of features.
So, our approach for inferring data precondition from a large DNN model with many 
layers is scalable because of quadratic time complexity. 

\section{Evaluation}
\label{sec:evaluation}
This section describes the evaluation of \deepinfer. First, we discuss the experimental setup in ~\secref{sec:experimentSetup}. Next, we describe research questions and present the results and discussion in~\secref{sec:results}.

\subsection{Experiment}
\label{sec:experimentSetup}

\subsubsection{Benchmark}
\label{sec:dataset}
We have gathered four canonical real-world datasets from Kaggle competitions~\cite{kaggle}. 
The train and test datasets are converted to numerical values if those are in any other data types during the data preprocessing stage. 
We have gathered models intended for classification problems from the Kaggle and used by prior work~\cite{udeshi,zhang,aggarwal,biswas2022fairify}.
In table~\ref{tab:datasetStat}, we present the total number of features in a dataset,  number of neurons, and layers of the models.

\subsubsection{Prediction interval}
\label{sec:pred}

\begin{table}[htbp]
\tiny
\centering
\caption{DNN Benchmark for inferring data preconditions}
\resizebox{\columnwidth}{!}{%
\begin{tabular}{|l|r|l|l|r|r|}
\hline
\rowcolor[HTML]{EFEFEF} 
\textbf{Dataset} &
  \textbf{\# Features} &
  \textbf{Model} &
  \textbf{Source} &
  \textbf{\# Layers} &
  \textbf{\# Neurons} \\ \hline
                         &                          & PD1  & \cellcolor[HTML]{FFFFFF}Kaggle                 & 3 & 221  \\ \cline{3-6} 
                         &                          & PD2  & \cellcolor[HTML]{FFFFFF}Kaggle                & 3 & 221  \\ \cline{3-6} 
                         &                          & PD3  & \cellcolor[HTML]{FFFFFF}Kaggle                & 3 & 221  \\ \cline{3-6} 
\multirow{-4}{*}{\textbf{Pima Diabetes}~\cite{pima} } &
  \multirow{-4}{*}{8} &
  PD4 &
  \cellcolor[HTML]{FFFFFF}Kaggle &
  4 &
  293 \\ \hline
\rowcolor[HTML]{F3F3F3} 
\cellcolor[HTML]{F3F3F3} & \cellcolor[HTML]{F3F3F3} & HP1  & Kaggle                                        & 3 & 273  \\ 
\rowcolor[HTML]{F3F3F3} 
\cline{3-6} 
\cellcolor[HTML]{F3F3F3} & \cellcolor[HTML]{F3F3F3} & HP2  & Kaggle                                        & 3 & 273  \\ 
\rowcolor[HTML]{F3F3F3} 
\cline{3-6} 
\cellcolor[HTML]{F3F3F3} & \cellcolor[HTML]{F3F3F3} & HP3  & Kaggle                                        & 3 & 273  \\ 
\rowcolor[HTML]{F3F3F3} 
\cline{3-6} 
\multirow{-4}{*}{\cellcolor[HTML]{F3F3F3}\textbf{House Price}~\cite{house} } &
  \multirow{-4}{*}{\cellcolor[HTML]{F3F3F3}10} &
  HP4 &
  Kaggle &
  4 &
  383 \\ \hline
                         &                          & BM1  & \cellcolor[HTML]{FFFFFF}{\cite{biswas2022fairify}}   & 4 & 97   \\ \cline{3-6} 
                         &                          & BM2  & \cellcolor[HTML]{FFFFFF}{\cite{biswas2022fairify}}   & 4 & 65   \\ \cline{3-6} 
                         &                          & BM3  & \cellcolor[HTML]{FFFFFF}{\cite{aggarwal}} & 3 & 117  \\ \cline{3-6} 
                         &                          & BM4  & \cellcolor[HTML]{FFFFFF}{\cite{biswas2022fairify}}   & 5 & 318  \\ \cline{3-6} 
                         &                          & BM5  & \cellcolor[HTML]{FFFFFF}{\cite{biswas2022fairify}}   & 4 & 49   \\ \cline{3-6} 
                         &                          & BM6  & \cellcolor[HTML]{FFFFFF}{\cite{biswas2022fairify}}   & 4 & 35   \\ \cline{3-6} 
                         &                          & BM7  & \cellcolor[HTML]{FFFFFF}{\cite{biswas2022fairify}}   & 4 & 145  \\ \cline{3-6} 
                         &                          & BM8  & \cellcolor[HTML]{FFFFFF}{\cite{zhang}}    & 7 & 141  \\ \cline{3-6} 
                         &                          & BM9  & \cellcolor[HTML]{FFFFFF}Kaggle                & 3 & 627  \\ \cline{3-6} 
                         &                          & BM10 & \cellcolor[HTML]{FFFFFF}Kaggle                & 3 & 627  \\ \cline{3-6} 
                         &                          & BM11 & \cellcolor[HTML]{FFFFFF}Kaggle                & 3 & 627  \\ \cline{3-6} 
\multirow{-12}{*}{\textbf{BankCustomer}~\cite{bank} } &
  \multirow{-12}{*}{28} &
  BM12 &
  \cellcolor[HTML]{FFFFFF}Kaggle &
  4 &
  1439 \\ \hline
\rowcolor[HTML]{F3F3F3} 
\cellcolor[HTML]{F3F3F3} & \cellcolor[HTML]{F3F3F3} & GC1  & {\cite{biswas2022fairify}}                           & 3 & 64   \\ \cline{3-6} 
\rowcolor[HTML]{F3F3F3} 
\cellcolor[HTML]{F3F3F3} & \cellcolor[HTML]{F3F3F3} & GC2  & {\cite{udeshi}}                           & 3 & 114  \\ \cline{3-6} 
\rowcolor[HTML]{F3F3F3} 
\cellcolor[HTML]{F3F3F3} & \cellcolor[HTML]{F3F3F3} & GC3  & {\cite{biswas2022fairify}}                           & 3 & 23   \\ \cline{3-6} 
\rowcolor[HTML]{F3F3F3} 
\cellcolor[HTML]{F3F3F3} & \cellcolor[HTML]{F3F3F3} & GC4  & {\cite{biswas2022fairify}}                           & 4 & 24   \\ \cline{3-6} 
\rowcolor[HTML]{F3F3F3} 
\cellcolor[HTML]{F3F3F3} & \cellcolor[HTML]{F3F3F3} & GC5  & {\cite{zhang}}                            & 7 & 138  \\ \cline{3-6} 
\rowcolor[HTML]{F3F3F3} 
\cellcolor[HTML]{F3F3F3} & \cellcolor[HTML]{F3F3F3} & GC6  & Kaggle                                        & 3 & 2397 \\ \cline{3-6} 
\rowcolor[HTML]{F3F3F3} 
\cellcolor[HTML]{F3F3F3} & \cellcolor[HTML]{F3F3F3} & GC7  & Kaggle                                        & 3 & 2397 \\ \cline{3-6} 
\rowcolor[HTML]{F3F3F3} 
\cellcolor[HTML]{F3F3F3} & \cellcolor[HTML]{F3F3F3} & GC8  & Kaggle                                        & 3 & 2397 \\ \cline{3-6} 
\rowcolor[HTML]{F3F3F3} 
\multirow{-9}{*}{\cellcolor[HTML]{F3F3F3}\textbf{GermanCredit}~\cite{germanCredit}} &
  \multirow{-9}{*}{\cellcolor[HTML]{F3F3F3}22} &
  GC9 &
  Kaggle &
  4 &
  2949 \\ \hline
\end{tabular}%
\label{tab:datasetStat}
}
\end{table}

We have adopted high-quality prediction intervals for deep learning models for classification and regression models from prior work~\cite{confidenceRef}. Therefore, for the experimental evaluation, we selected a prediction interval ($\ge 0.95$) as the postcondition for determining the data precondition from a deep learning model.  
\begin{table*}[htbp]
\setlength{\belowcaptionskip}{.05cm}
\caption{\deepinfer implying correct and incorrect model prediction for unseen data}
\footnotesize
\resizebox{\textwidth}{!}{%
\label{tab:rq1}
\begin{tabular}{|l|l|r|r|r|rr|rrrrr|r|}
\hline
\rowcolor[HTML]{CCCCCC} 
\cellcolor[HTML]{CCCCCC} &
  \cellcolor[HTML]{CCCCCC} &
  \cellcolor[HTML]{CCCCCC} &
  \cellcolor[HTML]{CCCCCC} &
  \cellcolor[HTML]{CCCCCC} &
  \multicolumn{2}{c|}{\cellcolor[HTML]{CCCCCC}\textbf{Ground Truth}} &
  \multicolumn{5}{c|}{\cellcolor[HTML]{CCCCCC}\textbf{DeepInfer}} &
  \cellcolor[HTML]{CCCCCC} \\ \cline{6-12}
\rowcolor[HTML]{CCCCCC} 
\multirow{-2}{*}{\cellcolor[HTML]{CCCCCC}\textbf{Dataset}} &
  \multirow{-2}{*}{\cellcolor[HTML]{CCCCCC}\textbf{Model}} &
  \multirow{-2}{*}{\cellcolor[HTML]{CCCCCC}\textbf{Accuracy}} &
  \multirow{-2}{*}{\cellcolor[HTML]{CCCCCC}\textbf{\# Features}} &
  \multirow{-2}{*}{\cellcolor[HTML]{CCCCCC}\textbf{\# Unseen data}} &
  \multicolumn{1}{r|}{\cellcolor[HTML]{CCCCCC}\textbf{\# Correct}} &
  \textbf{\# Incorrect} &
  \multicolumn{1}{r|}{\cellcolor[HTML]{CCCCCC}\textbf{\# Violation}} &
  \multicolumn{1}{r|}{\cellcolor[HTML]{CCCCCC}\textbf{\# Satisfaction}} &
  \multicolumn{1}{r|}{\cellcolor[HTML]{CCCCCC}\textbf{\# Correct}} &
  \multicolumn{1}{r|}{\cellcolor[HTML]{CCCCCC}\textbf{\# Incorrect}} &
  \textbf{\#Uncertain} &
  \multirow{-2}{*}{\cellcolor[HTML]{CCCCCC}\textbf{Time (sec)}} \\ \hline
 &
  \cellcolor[HTML]{D9EAD3}PD1 &
  \cellcolor[HTML]{D9EAD3}77.98\% &
   &
   &
  \multicolumn{1}{r|}{\cellcolor[HTML]{D9EAD3}119} &
  \cellcolor[HTML]{D9EAD3}34 &
  \multicolumn{1}{r|}{\cellcolor[HTML]{D9EAD3}192} &
  \multicolumn{1}{r|}{\cellcolor[HTML]{D9EAD3}1032} &
  \multicolumn{1}{r|}{\cellcolor[HTML]{D9EAD3}108} &
  \multicolumn{1}{r|}{\cellcolor[HTML]{D9EAD3}43} &
  \cellcolor[HTML]{D9EAD3}2 &
  0.67 \\ \cline{2-3} \cline{6-13} 
 &
  PD2 &
  65.10\% &
   &
   &
  \multicolumn{1}{r|}{99} &
  54 &
  \multicolumn{1}{r|}{0} &
  \multicolumn{1}{r|}{1224} &
  \multicolumn{1}{r|}{153} &
  \multicolumn{1}{r|}{0} &
  0 &
  0.66 \\ \cline{2-3} \cline{6-13} 
 &
  PD3 &
  65.49\% &
   &
   &
  \multicolumn{1}{r|}{\cellcolor[HTML]{FFFFFF}98} &
  \cellcolor[HTML]{FFFFFF}55 &
  \multicolumn{1}{r|}{129} &
  \multicolumn{1}{r|}{1095} &
  \multicolumn{1}{r|}{\cellcolor[HTML]{FFFFFF}74} &
  \multicolumn{1}{r|}{\cellcolor[HTML]{FFFFFF}79} &
  \cellcolor[HTML]{FFFFFF}0 &
  0.65 \\ \cline{2-3} \cline{6-13} 
\multirow{-4}{*}{\textbf{Pima Diabetes}} &
  PD4 &
  77.47\% &
  \multirow{-4}{*}{8} &
  \multirow{-4}{*}{153} &
  \multicolumn{1}{r|}{111} &
  42 &
  \multicolumn{1}{r|}{132} &
  \multicolumn{1}{r|}{1092} &
  \multicolumn{1}{r|}{37} &
  \multicolumn{1}{r|}{116} &
  0 &
  0.65 \\ \hline
\rowcolor[HTML]{F3F3F3} 
\cellcolor[HTML]{F3F3F3} &
  HP1 &
  85.22\% &
  \cellcolor[HTML]{F3F3F3} &
  \cellcolor[HTML]{F3F3F3} &
  \multicolumn{1}{r|}{\cellcolor[HTML]{F3F3F3}147} &
  145 &
  \multicolumn{1}{r|}{\cellcolor[HTML]{F3F3F3}341} &
  \multicolumn{1}{r|}{\cellcolor[HTML]{F3F3F3}2579} &
  \multicolumn{1}{r|}{\cellcolor[HTML]{F3F3F3}188} &
  \multicolumn{1}{r|}{\cellcolor[HTML]{F3F3F3}98} &
  6 &
  0.86 \\ \cline{2-3} \cline{6-13} 
\rowcolor[HTML]{D9EAD3} 
\cellcolor[HTML]{F3F3F3} &
  HP2 &
  89.77\% &
  \cellcolor[HTML]{F3F3F3} &
  \cellcolor[HTML]{F3F3F3} &
  \multicolumn{1}{r|}{\cellcolor[HTML]{D9EAD3}147} &
  145 &
  \multicolumn{1}{r|}{\cellcolor[HTML]{D9EAD3}341} &
  \multicolumn{1}{r|}{\cellcolor[HTML]{D9EAD3}2579} &
  \multicolumn{1}{r|}{\cellcolor[HTML]{D9EAD3}188} &
  \multicolumn{1}{r|}{\cellcolor[HTML]{D9EAD3}98} &
  6 &
  \cellcolor[HTML]{EFEFEF}0.83 \\ \cline{2-3} \cline{6-13} 
\rowcolor[HTML]{F3F3F3} 
\cellcolor[HTML]{F3F3F3} &
  HP3 &
  45.45\% &
  \cellcolor[HTML]{F3F3F3} &
  \cellcolor[HTML]{F3F3F3} &
  \multicolumn{1}{r|}{\cellcolor[HTML]{F3F3F3}145} &
  147 &
  \multicolumn{1}{r|}{\cellcolor[HTML]{F3F3F3}0} &
  \multicolumn{1}{r|}{\cellcolor[HTML]{F3F3F3}2920} &
  \multicolumn{1}{r|}{\cellcolor[HTML]{F3F3F3}292} &
  \multicolumn{1}{r|}{\cellcolor[HTML]{F3F3F3}0} &
  0 &
  0.97 \\ \cline{2-3} \cline{6-13} 
\rowcolor[HTML]{F3F3F3} 
\multirow{-4}{*}{\cellcolor[HTML]{F3F3F3}\textbf{House Price}} &
  HP4 &
  87.50\% &
  \multirow{-4}{*}{\cellcolor[HTML]{F3F3F3}10} &
  \multirow{-4}{*}{\cellcolor[HTML]{F3F3F3}292} &
  \multicolumn{1}{r|}{\cellcolor[HTML]{F3F3F3}147} &
  145 &
  \multicolumn{1}{r|}{\cellcolor[HTML]{F3F3F3}188} &
  \multicolumn{1}{r|}{\cellcolor[HTML]{F3F3F3}2732} &
  \multicolumn{1}{r|}{\cellcolor[HTML]{F3F3F3}107} &
  \multicolumn{1}{r|}{\cellcolor[HTML]{F3F3F3}184} &
  1 &
  0.87 \\ \hline
 &
  BM1 &
  81.10\% &
   &
   &
  \multicolumn{1}{r|}{1072} &
  1044 &
  \multicolumn{1}{r|}{18814} &
  \multicolumn{1}{r|}{40434} &
  \multicolumn{1}{r|}{616} &
  \multicolumn{1}{r|}{1500} &
  0 &
  3.42 \\ \cline{2-3} \cline{6-13} 
 &
  BM2 &
  82.11\% &
   &
   &
  \multicolumn{1}{r|}{1066} &
  1050 &
  \multicolumn{1}{r|}{14855} &
  \multicolumn{1}{r|}{44393} &
  \multicolumn{1}{r|}{1492} &
  \multicolumn{1}{r|}{624} &
  0 &
  3.48 \\ \cline{2-3} \cline{6-13} 
 &
  BM3 &
  80.19\% &
   &
   &
  \multicolumn{1}{r|}{\cellcolor[HTML]{FFFFFF}1054} &
  \cellcolor[HTML]{FFFFFF}1062 &
  \multicolumn{1}{r|}{7370} &
  \multicolumn{1}{r|}{51878} &
  \multicolumn{1}{r|}{\cellcolor[HTML]{FFFFFF}734} &
  \multicolumn{1}{r|}{\cellcolor[HTML]{FFFFFF}1382} &
  \cellcolor[HTML]{FFFFFF}0 &
  3.78 \\ \cline{2-3} \cline{6-13} 
 &
  BM4 &
  79.10\% &
   &
   &
  \multicolumn{1}{r|}{1067} &
  1049 &
  \multicolumn{1}{r|}{17486} &
  \multicolumn{1}{r|}{41762} &
  \multicolumn{1}{r|}{1061} &
  \multicolumn{1}{r|}{1055} &
  0 &
  3.78 \\ \cline{2-3} \cline{6-13} 
 &
  BM5 &
  82.10\% &
   &
   &
  \multicolumn{1}{r|}{1052} &
  1064 &
  \multicolumn{1}{r|}{7703} &
  \multicolumn{1}{r|}{51545} &
  \multicolumn{1}{r|}{807} &
  \multicolumn{1}{r|}{1306} &
  3 &
  3.39 \\ \cline{2-3} \cline{6-13} 
 &
  BM6 &
  82.00\% &
   &
   &
  \multicolumn{1}{r|}{1059} &
  1057 &
  \multicolumn{1}{r|}{17868} &
  \multicolumn{1}{r|}{41380} &
  \multicolumn{1}{r|}{1099} &
  \multicolumn{1}{r|}{1017} &
  0 &
  3.48 \\ \cline{2-3} \cline{6-13} 
 &
  BM7 &
  81.00\% &
   &
   &
  \multicolumn{1}{r|}{1074} &
  1042 &
  \multicolumn{1}{r|}{12762} &
  \multicolumn{1}{r|}{46486} &
  \multicolumn{1}{r|}{1375} &
  \multicolumn{1}{r|}{741} &
  0 &
  3.40 \\ \cline{2-3} \cline{6-13} 
 &
  BM8 &
  82.00\% &
   &
   &
  \multicolumn{1}{r|}{1075} &
  1041 &
  \multicolumn{1}{r|}{15213} &
  \multicolumn{1}{r|}{44035} &
  \multicolumn{1}{r|}{1089} &
  \multicolumn{1}{r|}{1027} &
  0 &
  3.90 \\ \cline{2-3} \cline{6-13} 
 &
  BM9 &
  81.30\% &
   &
   &
  \multicolumn{1}{r|}{1058} &
  1058 &
  \multicolumn{1}{r|}{21395} &
  \multicolumn{1}{r|}{37853} &
  \multicolumn{1}{r|}{1392} &
  \multicolumn{1}{r|}{724} &
  0 &
  3.32 \\ \cline{2-3} \cline{6-13} 
 &
  BM10 &
  81.90\% &
   &
   &
  \multicolumn{1}{r|}{1092} &
  1024 &
  \multicolumn{1}{r|}{9931} &
  \multicolumn{1}{r|}{49317} &
  \multicolumn{1}{r|}{820} &
  \multicolumn{1}{r|}{1296} &
  0 &
  3.21 \\ \cline{2-3} \cline{6-13} 
 &
  \cellcolor[HTML]{D9EAD3}BM11 &
  \cellcolor[HTML]{D9EAD3}83.60\% &
   &
   &
  \multicolumn{1}{r|}{\cellcolor[HTML]{D9EAD3}1092} &
  \cellcolor[HTML]{D9EAD3}1024 &
  \multicolumn{1}{r|}{\cellcolor[HTML]{D9EAD3}27241} &
  \multicolumn{1}{r|}{\cellcolor[HTML]{D9EAD3}32007} &
  \multicolumn{1}{r|}{\cellcolor[HTML]{D9EAD3}945} &
  \multicolumn{1}{r|}{\cellcolor[HTML]{D9EAD3}1171} &
  \cellcolor[HTML]{D9EAD3}0 &
  \cellcolor[HTML]{FFFFFF}3.10 \\ \cline{2-3} \cline{6-13} 
\multirow{-12}{*}{\textbf{BankCustomer}} &
  BM12 &
  80.70\% &
  \multirow{-12}{*}{28} &
  \multirow{-12}{*}{2116} &
  \multicolumn{1}{r|}{1075} &
  1041 &
  \multicolumn{1}{r|}{20051} &
  \multicolumn{1}{r|}{39197} &
  \multicolumn{1}{r|}{1164} &
  \multicolumn{1}{r|}{952} &
  0 &
  3.23 \\ \hline
\rowcolor[HTML]{D9EAD3} 
\cellcolor[HTML]{F3F3F3} &
  GC1 &
  99.00\% &
  \cellcolor[HTML]{F3F3F3} &
  \cellcolor[HTML]{F3F3F3} &
  \multicolumn{1}{r|}{\cellcolor[HTML]{D9EAD3}198} &
  2 &
  \multicolumn{1}{r|}{\cellcolor[HTML]{D9EAD3}1044} &
  \multicolumn{1}{r|}{\cellcolor[HTML]{D9EAD3}3356} &
  \multicolumn{1}{r|}{\cellcolor[HTML]{D9EAD3}200} &
  \multicolumn{1}{r|}{\cellcolor[HTML]{D9EAD3}0} &
  0 &
  \cellcolor[HTML]{EFEFEF}1.94 \\ \cline{2-3} \cline{6-13} 
\rowcolor[HTML]{F3F3F3} 
\cellcolor[HTML]{F3F3F3} &
  GC2 &
  99.00\% &
  \cellcolor[HTML]{F3F3F3} &
  \cellcolor[HTML]{F3F3F3} &
  \multicolumn{1}{r|}{\cellcolor[HTML]{F3F3F3}198} &
  2 &
  \multicolumn{1}{r|}{\cellcolor[HTML]{F3F3F3}959} &
  \multicolumn{1}{r|}{\cellcolor[HTML]{F3F3F3}3441} &
  \multicolumn{1}{r|}{\cellcolor[HTML]{F3F3F3}188} &
  \multicolumn{1}{r|}{\cellcolor[HTML]{F3F3F3}12} &
  0 &
  2.18 \\ \cline{2-3} \cline{6-13} 
\rowcolor[HTML]{F3F3F3} 
\cellcolor[HTML]{F3F3F3} &
  GC3 &
  99.00\% &
  \cellcolor[HTML]{F3F3F3} &
  \cellcolor[HTML]{F3F3F3} &
  \multicolumn{1}{r|}{\cellcolor[HTML]{F3F3F3}198} &
  2 &
  \multicolumn{1}{r|}{\cellcolor[HTML]{F3F3F3}1569} &
  \multicolumn{1}{r|}{\cellcolor[HTML]{F3F3F3}2831} &
  \multicolumn{1}{r|}{\cellcolor[HTML]{F3F3F3}73} &
  \multicolumn{1}{r|}{\cellcolor[HTML]{F3F3F3}127} &
  0 &
  2.03 \\ \cline{2-3} \cline{6-13} 
\rowcolor[HTML]{D9EAD3} 
\cellcolor[HTML]{F3F3F3} &
  GC4 &
  99.00\% &
  \cellcolor[HTML]{F3F3F3} &
  \cellcolor[HTML]{F3F3F3} &
  \multicolumn{1}{r|}{\cellcolor[HTML]{D9EAD3}198} &
  2 &
  \multicolumn{1}{r|}{\cellcolor[HTML]{D9EAD3}2401} &
  \multicolumn{1}{r|}{\cellcolor[HTML]{D9EAD3}1999} &
  \multicolumn{1}{r|}{\cellcolor[HTML]{D9EAD3}200} &
  \multicolumn{1}{r|}{\cellcolor[HTML]{D9EAD3}0} &
  0 &
  \cellcolor[HTML]{EFEFEF}2.01 \\ \cline{2-3} \cline{6-13} 
\rowcolor[HTML]{F3F3F3} 
\cellcolor[HTML]{F3F3F3} &
  GC5 &
  99.00\% &
  \cellcolor[HTML]{F3F3F3} &
  \cellcolor[HTML]{F3F3F3} &
  \multicolumn{1}{r|}{\cellcolor[HTML]{F3F3F3}198} &
  2 &
  \multicolumn{1}{r|}{\cellcolor[HTML]{F3F3F3}1193} &
  \multicolumn{1}{r|}{\cellcolor[HTML]{F3F3F3}3207} &
  \multicolumn{1}{r|}{\cellcolor[HTML]{F3F3F3}195} &
  \multicolumn{1}{r|}{\cellcolor[HTML]{F3F3F3}5} &
  0 &
  1.93 \\ \cline{2-3} \cline{6-13} 
\rowcolor[HTML]{F3F3F3} 
\cellcolor[HTML]{F3F3F3} &
  GC6 &
  99.00\% &
  \cellcolor[HTML]{F3F3F3} &
  \cellcolor[HTML]{F3F3F3} &
  \multicolumn{1}{r|}{\cellcolor[HTML]{F3F3F3}198} &
  2 &
  \multicolumn{1}{r|}{\cellcolor[HTML]{F3F3F3}1627} &
  \multicolumn{1}{r|}{\cellcolor[HTML]{F3F3F3}2773} &
  \multicolumn{1}{r|}{\cellcolor[HTML]{F3F3F3}67} &
  \multicolumn{1}{r|}{\cellcolor[HTML]{F3F3F3}133} &
  0 &
  1.99 \\ \cline{2-3} \cline{6-13} 
\rowcolor[HTML]{F3F3F3} 
\cellcolor[HTML]{F3F3F3} &
  GC7 &
  99.00\% &
  \cellcolor[HTML]{F3F3F3} &
  \cellcolor[HTML]{F3F3F3} &
  \multicolumn{1}{r|}{\cellcolor[HTML]{F3F3F3}198} &
  2 &
  \multicolumn{1}{r|}{\cellcolor[HTML]{F3F3F3}1074} &
  \multicolumn{1}{r|}{\cellcolor[HTML]{F3F3F3}3326} &
  \multicolumn{1}{r|}{\cellcolor[HTML]{F3F3F3}195} &
  \multicolumn{1}{r|}{\cellcolor[HTML]{F3F3F3}5} &
  0 &
  1.96 \\ \cline{2-3} \cline{6-13} 
\rowcolor[HTML]{F3F3F3} 
\cellcolor[HTML]{F3F3F3} &
  GC8 &
  99.00\% &
  \cellcolor[HTML]{F3F3F3} &
  \cellcolor[HTML]{F3F3F3} &
  \multicolumn{1}{r|}{\cellcolor[HTML]{F3F3F3}198} &
  2 &
  \multicolumn{1}{r|}{\cellcolor[HTML]{F3F3F3}1360} &
  \multicolumn{1}{r|}{\cellcolor[HTML]{F3F3F3}3040} &
  \multicolumn{1}{r|}{\cellcolor[HTML]{F3F3F3}143} &
  \multicolumn{1}{r|}{\cellcolor[HTML]{F3F3F3}57} &
  0 &
  2.07 \\ \cline{2-3} \cline{6-13} 
\rowcolor[HTML]{F3F3F3} 
\multirow{-9}{*}{\cellcolor[HTML]{F3F3F3}\textbf{GermanCredit}} &
  GC9 &
  99.00\% &
  \multirow{-9}{*}{\cellcolor[HTML]{F3F3F3}22} &
  \multirow{-9}{*}{\cellcolor[HTML]{F3F3F3}200} &
  \multicolumn{1}{r|}{\cellcolor[HTML]{F3F3F3}198} &
  2 &
  \multicolumn{1}{r|}{\cellcolor[HTML]{F3F3F3}1360} &
  \multicolumn{1}{r|}{\cellcolor[HTML]{F3F3F3}3040} &
  \multicolumn{1}{r|}{\cellcolor[HTML]{F3F3F3}143} &
  \multicolumn{1}{r|}{\cellcolor[HTML]{F3F3F3}57} &
  0 &
  1.93 \\ \hline
\end{tabular}%
}
\end{table*}

\subsubsection{Experimental Setup}
\label{sec:implementation}
To perform our experiments and evaluation, we implemented our techniques using \textit{Python} and \textit{Keras}. We have used mathematical packages (\textit{numpy}, \textit{pandas}) to compute the data precondition from a \textit{Keras} model and to evaluate the implied trustworthiness of model’s prediction using inferred data preconditions. We have conducted all the experiments on a machine with a 2 GHz Quad-Core Intel Core i7 and 32 GB 1867 MHz DDR3 RAM running the macOS 11.14.

\subsubsection{Evaluation Metrics}:
\label{sec:metric}
To determine the efficiency \deepinfer, we measure the Pearson Correlation Coefficient $(pcc)$  following prior work~\cite{fariha2021conformance}. We define true positive (TP), false positive (FP), false negative (FN), and true negative (TN) following prior work~\cite{selfchecker}. We also measure precision, recall, TPR, FPR, F-1 score following prior work~\cite{selfchecker} from TP, FP, and FN to determine the efficiency of our approach in predicting the correct prediction of a DNN model.
\subsection{Results}
\label{sec:results}
\subsubsection{Research Questions}
\label{sec:rq}
To evaluate the utility, efficiency, and performance, we answer the following research questions:

\textbf{RQ1(Utility):} \textit{Do data precondition violations imply incorrect model prediction, and data precondition satisfaction implies correct model prediction, i.e., to trust the model? }

We first obtain the preconditions on data for each feature using the respective model and dataset to measure the utility of data for implying the model's prediction. Then using Algorithm~\ref{algoUnseen2}, we imply "Correct" or "Incorrect" or  "Uncertain" prediction for unseen data based on data precondition violation and satisfaction for each feature. For RQ1, the model has been trained with the seen i.e., training data, and validated with the second portion of training data. Following the experimental procedure~\cite{selfchecker}, we have used all the test datasets as unseen data. 
For evaluation purposes, we determine the ground truth from the actual label and the model's predicted label and we consider "Uncertain" prediction as "Incorrect". 

\textbf{RQ2 (Effectiveness):} \textit{How effective \deepinfer is to imply trustworthiness in the model's prediction compared to the prior approach? }

To determine the effectiveness of our proposed approach \deepinfer, we measure true positive, false positive, false negative, and true negative as discussed in ~\secref{sec:metric} . 
We reported the false positive and true positive ground truth where "ActFP" denotes if the actual label and predicted label by a model are not equal and "ActTP" denotes if the actual label and predicted label by a model are equal. This suggests whether the model is properly trained or not and also explains how \deepinfer performs compared to the "ActFP" and "ActTP". We compare our approach with \selfchecker~\cite{selfchecker} using same 29 models and 4 datasets. We have compared our approach against \selfchecker~\cite{selfchecker} in terms of how effective each approach is in predicting DNN misclassifications in deployment. We have used the open-source implementation of \selfchecker and utilized the same  hardware setup. We communicated with the authors to ensure their tool is applicable to these models and datasets. 

\textbf{RQ3 (Efficiency):} \textit{ What is the performance of \deepinfer with respect to time, and what is the runtime overhead using unseen data during deployment compared to prior work?} 

To compute the efficiency of our proposed technique, we compute the training time of all the models. We computed the runtime of \deepinfer and \selfchecker for all the models and all unseen datasets. 
We consider the runtime measure important for determining trust on the model’s prediction with unseen data in the deployment stage for safety-critical issues. 
Considering resource constraints such as processing data and generating prediction timely and limited computing power or memory, it is crucial to ensure that models are suitable for deployment in safety-critical scenarios to prevent accidents or mitigate risks. For instance, a self-driving Uber car struck and killed a woman in March 2018 as an investigation~\cite{Uber} revealed that the model couldn't correctly predict her path and it needed to brake just 1.3 seconds before it struck her. Therefore, it is important to measure the runtime of such techniques.

\subsubsection{Results and Analysis}
\label{sec:resultsana}
In this section, we discuss the results and analysis for each of the research questions utilizing 4 different real-world tabular datasets with 29 different \keras real-world models (discussed in~\secref{sec:dataset})
targeting binary classification problems.
 
\textbf{RQ1 (Utility):} 
For RQ1, we present the results of all 29 real-world models for four different datasets in Table~\ref{tab:rq1}. We report the model's accuracy and the number of test instances. Then, we reported the total number of "Correct" and "Incorrect" labels for all the test datasets as the ground truth of the model's prediction and actual label. Next, we report the total number of data precondition violations and satisfaction. Then, we report "Correct" and "Incorrect" implications in "\#Correct" and "\#Incorrect", "\#Unseen" columns using our proposed technique \deepinfer. We also measure the total runtime and report in the "Time" column in Table~\ref{tab:rq1}. 
From the results, we observe that for the model with high accuracy, the total number of "Correct" and "Incorrect" implied using \deepinfer is comparable to the ground truth. For example, for the German Credit dataset and GC1 and GC4 model with accuracy 99.00\%, \deepinfer obtained 200 "\#Correct" and 0 "\#Incorrect" where Ground Truth contains in total 198 "\#Correct" and 2 "\#Incorrect" labels. 
The reason behind incorrectly implying a number of incorrect and correct predictions in models like BM11 is that the model itself was not trained well, as low accuracy suggests. 
Based on our findings, we conclude that the model with high accuracy implies a better comparable number of "Correct" and "Incorrect" predictions for all the unseen datasets. Despite several models exhibiting high accuracy, we observed a lack of correlation between the number of violations and the accuracy of these models. This finding suggests the presence of underlying issues that warrant further investigation. 
We investigated further to determine the correlation between the number of violations in data preconditions and the frequency of "Correct" and "Incorrect" predictions based on the ground truth. Using the Pearson Correlation Coefficient (pcc) following prior work~\cite{fariha2021conformance}, we found a positive correlation of 0.88 between data precondition violations and incorrect model predictions, indicating that as the number of violations increases, the likelihood of incorrect predictions by the model also rises. This highlights the importance of data preconditions in determining the trustworthiness of the model's predictions. Additionally, we saw a strong correlation of 0.98 between precondition satisfaction and correct model predictions, indicating that the model tends to make accurate predictions when data preconditions are satisfied.
To assess the statistical significance of these correlations, we conducted a t-test to compute p-values following prior work~\cite{fariha2021conformance}, yielding p-values of 0.0001 for the correlation between data precondition violation and incorrect prediction and 0.0003 for the correlation between data precondition satisfaction and correct prediction.  Based on the commonly used significance level of 0.05, these p-values indicate that the correlations are statistically significant~\cite{rice1989analyzing}. A p-value below 0.05 suggests strong evidence against the null hypothesis, supporting the presence of a significant correlation between the variables.

\textit{ In summary, \deepinfer implies that data precondition violations and Incorrect model prediction are highly correlated (0.88) between prediction ground truth and violation. Also, the precondition satisfaction and correct model prediction are strongly correlated (0.98).} 

\textbf{RQ2 (Effectiveness):}  
In Table~\ref{tab:rq2}, we highlighted the best values with high model accuracy from each set of the dataset. We also observe how close the values are obtained from \deepinfer compared to the ground truth FP and TP. 
Some of the models, e.g., BM6, BM7, BM9, BM10, BM11 in the Bank Customer dataset throw \textit{numpy.linalg.LinAlgError:Singular matrix} error during KDE generation steps using \selfchecker tool. We communicated with the authors of \selfchecker, and they explained that 
the models they used for evaluation contained only \texttt{relu,softmax} having more than 8 layers for image datasets. Furthermore, we obtain 0 FP and 0 TP and the same number of FN and TN for many models under experiments e.g., PD2, PD3, HP2, Hp3, BM4, BM5, BM8, GC5, GC6, GC7, GC8, and GC9 etc. We have investigated further and found that \selfchecker approach does not handle a model if the last layer contains \texttt{sigmoid, relu, tanh} activation functions with single output and the threshold of KDE values performs well for \texttt{softmax} activation functions with multiple outputs to determine true misbehavior of the model. 
\begin{table*}[htbp]
\setlength{\belowcaptionskip}{.05cm}
\caption{Efficiency of \deepinfer for implying model’s prediction}
\footnotesize
\resizebox{\textwidth}{!}{%
\begin{tabular}{|l|l|r|rr|rrrrrrrrrr|rrrrrrrrrr|}
\hline
\cellcolor[HTML]{D9D9D9} &
  \cellcolor[HTML]{D9D9D9} &
  \cellcolor[HTML]{D9D9D9} &
  \multicolumn{2}{c|}{\cellcolor[HTML]{D9D9D9}\textbf{Ground Truth}} &
  \multicolumn{10}{c|}{\cellcolor[HTML]{D0E0E3}\textbf{SelfChecker}} &
  \multicolumn{10}{c|}{\cellcolor[HTML]{D9D2E9}\textbf{DeepInfer}} \\ \cline{4-25} 
\multirow{-2}{*}{\cellcolor[HTML]{D9D9D9}\textbf{Dataset}} &
  \multirow{-2}{*}{\cellcolor[HTML]{D9D9D9}\textbf{Model}} &
  \multirow{-2}{*}{\cellcolor[HTML]{D9D9D9}\textbf{Test Acc.}} &
  \multicolumn{1}{r|}{\cellcolor[HTML]{D9D9D9}\textbf{ActFP}} &
  \cellcolor[HTML]{D9D9D9}\textbf{ActTP} &
  \multicolumn{1}{r|}{\cellcolor[HTML]{D0E0E3}\textbf{FP}} &
  \multicolumn{1}{r|}{\cellcolor[HTML]{D0E0E3}\textbf{TP}} &
  \multicolumn{1}{r|}{\cellcolor[HTML]{D0E0E3}\textbf{FN}} &
  \multicolumn{1}{r|}{\cellcolor[HTML]{D0E0E3}\textbf{TN}} &
  \multicolumn{1}{l|}{\cellcolor[HTML]{D0E0E3}\textbf{Precision}} &
  \multicolumn{1}{l|}{\cellcolor[HTML]{D0E0E3}\textbf{Recall}} &
  \multicolumn{1}{l|}{\cellcolor[HTML]{D0E0E3}\textbf{Accuracy}} &
  \multicolumn{1}{l|}{\cellcolor[HTML]{D0E0E3}\textbf{TPR}} &
  \multicolumn{1}{l|}{\cellcolor[HTML]{D0E0E3}\textbf{FPR}} &
  \multicolumn{1}{l|}{\cellcolor[HTML]{D0E0E3}\textbf{F-1}} &
  \multicolumn{1}{r|}{\cellcolor[HTML]{D9D2E9}\textbf{FP}} &
  \multicolumn{1}{r|}{\cellcolor[HTML]{D9D2E9}\textbf{TP}} &
  \multicolumn{1}{r|}{\cellcolor[HTML]{D9D2E9}\textbf{FN}} &
  \multicolumn{1}{r|}{\cellcolor[HTML]{D9D2E9}\textbf{TN}} &
  \multicolumn{1}{l|}{\cellcolor[HTML]{D9D2E9}\textbf{Precision}} &
  \multicolumn{1}{l|}{\cellcolor[HTML]{D9D2E9}\textbf{Recall}} &
  \multicolumn{1}{l|}{\cellcolor[HTML]{D9D2E9}\textbf{Accuracy}} &
  \multicolumn{1}{l|}{\cellcolor[HTML]{D9D2E9}\textbf{TPR}} &
  \multicolumn{1}{l|}{\cellcolor[HTML]{D9D2E9}\textbf{FPR}} &
  \multicolumn{1}{l|}{\cellcolor[HTML]{D9D2E9}\textbf{F-1}} \\ \hline
 &
  \cellcolor[HTML]{D9EAD3}PD1 &
  \cellcolor[HTML]{D9EAD3}77.98\% &
  \multicolumn{1}{r|}{34} &
  119 &
  \multicolumn{1}{r|}{46} &
  \multicolumn{1}{r|}{90} &
  \multicolumn{1}{r|}{5} &
  \multicolumn{1}{r|}{12} &
  \multicolumn{1}{r|}{0.66} &
  \multicolumn{1}{r|}{0.95} &
  \multicolumn{1}{r|}{0.67} &
  \multicolumn{1}{r|}{0.95} &
  \multicolumn{1}{r|}{0.79} &
  0.78 &
  \multicolumn{1}{r|}{33} &
  \multicolumn{1}{r|}{118} &
  \multicolumn{1}{r|}{1} &
  \multicolumn{1}{r|}{1} &
  \multicolumn{1}{r|}{\cellcolor[HTML]{D9EAD3}0.78} &
  \multicolumn{1}{r|}{\cellcolor[HTML]{D9EAD3}0.99} &
  \multicolumn{1}{r|}{\cellcolor[HTML]{D9EAD3}0.78} &
  \multicolumn{1}{r|}{\cellcolor[HTML]{D9EAD3}0.99} &
  \multicolumn{1}{r|}{\cellcolor[HTML]{D9EAD3}0.97} &
  \cellcolor[HTML]{D9EAD3}0.87 \\ \cline{2-25} 
 &
  PD2 &
  65.10\% &
  \multicolumn{1}{r|}{54} &
  99 &
  \multicolumn{1}{r|}{0} &
  \multicolumn{1}{r|}{0} &
  \multicolumn{1}{r|}{59} &
  \multicolumn{1}{r|}{94} &
  \multicolumn{1}{r|}{-} &
  \multicolumn{1}{r|}{0.00} &
  \multicolumn{1}{r|}{0.61} &
  \multicolumn{1}{r|}{0.00} &
  \multicolumn{1}{r|}{0.00} &
  0.00 &
  \multicolumn{1}{r|}{54} &
  \multicolumn{1}{r|}{99} &
  \multicolumn{1}{r|}{0} &
  \multicolumn{1}{r|}{0} &
  \multicolumn{1}{r|}{0.65} &
  \multicolumn{1}{r|}{1.00} &
  \multicolumn{1}{r|}{0.65} &
  \multicolumn{1}{r|}{1.00} &
  \multicolumn{1}{r|}{1.00} &
  0.79 \\ \cline{2-25} 
 &
  PD3 &
  65.49\% &
  \multicolumn{1}{r|}{55} &
  98 &
  \multicolumn{1}{r|}{0} &
  \multicolumn{1}{r|}{0} &
  \multicolumn{1}{r|}{59} &
  \multicolumn{1}{r|}{94} &
  \multicolumn{1}{r|}{-} &
  \multicolumn{1}{r|}{0.00} &
  \multicolumn{1}{r|}{0.61} &
  \multicolumn{1}{r|}{0.00} &
  \multicolumn{1}{r|}{0.00} &
  0.00 &
  \multicolumn{1}{r|}{55} &
  \multicolumn{1}{r|}{98} &
  \multicolumn{1}{r|}{0} &
  \multicolumn{1}{r|}{0} &
  \multicolumn{1}{r|}{0.64} &
  \multicolumn{1}{r|}{1.00} &
  \multicolumn{1}{r|}{0.64} &
  \multicolumn{1}{r|}{1.00} &
  \multicolumn{1}{r|}{1.00} &
  0.78 \\ \cline{2-25} 
\multirow{-4}{*}{Pima Diabetes} &
  PD4 &
  77.47\% &
  \multicolumn{1}{r|}{42} &
  111 &
  \multicolumn{1}{r|}{60} &
  \multicolumn{1}{r|}{77} &
  \multicolumn{1}{r|}{7} &
  \multicolumn{1}{r|}{9} &
  \multicolumn{1}{r|}{0.56} &
  \multicolumn{1}{r|}{0.92} &
  \multicolumn{1}{r|}{0.56} &
  \multicolumn{1}{r|}{0.92} &
  \multicolumn{1}{r|}{0.87} &
  0.70 &
  \multicolumn{1}{r|}{42} &
  \multicolumn{1}{r|}{111} &
  \multicolumn{1}{r|}{0} &
  \multicolumn{1}{r|}{0} &
  \multicolumn{1}{r|}{0.73} &
  \multicolumn{1}{r|}{1.00} &
  \multicolumn{1}{r|}{0.73} &
  \multicolumn{1}{r|}{1.00} &
  \multicolumn{1}{r|}{1.00} &
  0.84 \\ \hline
\rowcolor[HTML]{F3F3F3} 
\cellcolor[HTML]{F3F3F3} &
  HP1 &
  85.22\% &
  \multicolumn{1}{r|}{\cellcolor[HTML]{F3F3F3}145} &
  147 &
  \multicolumn{1}{r|}{\cellcolor[HTML]{F3F3F3}59} &
  \multicolumn{1}{r|}{\cellcolor[HTML]{F3F3F3}114} &
  \multicolumn{1}{r|}{\cellcolor[HTML]{F3F3F3}13} &
  \multicolumn{1}{r|}{\cellcolor[HTML]{F3F3F3}105} &
  \multicolumn{1}{r|}{\cellcolor[HTML]{F3F3F3}0.66} &
  \multicolumn{1}{r|}{\cellcolor[HTML]{F3F3F3}0.90} &
  \multicolumn{1}{r|}{\cellcolor[HTML]{F3F3F3}0.75} &
  \multicolumn{1}{r|}{\cellcolor[HTML]{F3F3F3}0.90} &
  \multicolumn{1}{r|}{\cellcolor[HTML]{F3F3F3}0.36} &
  0.76 &
  \multicolumn{1}{r|}{\cellcolor[HTML]{F3F3F3}127} &
  \multicolumn{1}{r|}{\cellcolor[HTML]{F3F3F3}145} &
  \multicolumn{1}{r|}{\cellcolor[HTML]{F3F3F3}2} &
  \multicolumn{1}{r|}{\cellcolor[HTML]{F3F3F3}18} &
  \multicolumn{1}{r|}{\cellcolor[HTML]{F3F3F3}0.53} &
  \multicolumn{1}{r|}{\cellcolor[HTML]{F3F3F3}0.99} &
  \multicolumn{1}{r|}{\cellcolor[HTML]{F3F3F3}0.56} &
  \multicolumn{1}{r|}{\cellcolor[HTML]{F3F3F3}0.99} &
  \multicolumn{1}{r|}{\cellcolor[HTML]{F3F3F3}0.88} &
  0.69 \\ \cline{2-25} 
\rowcolor[HTML]{F3F3F3} 
\cellcolor[HTML]{F3F3F3} &
  \cellcolor[HTML]{D9EAD3}HP2 &
  \cellcolor[HTML]{D9EAD3}89.77\% &
  \multicolumn{1}{r|}{\cellcolor[HTML]{F3F3F3}145} &
  147 &
  \multicolumn{1}{r|}{\cellcolor[HTML]{F3F3F3}0} &
  \multicolumn{1}{r|}{\cellcolor[HTML]{F3F3F3}0} &
  \multicolumn{1}{r|}{\cellcolor[HTML]{F3F3F3}139} &
  \multicolumn{1}{r|}{\cellcolor[HTML]{F3F3F3}153} &
  \multicolumn{1}{r|}{\cellcolor[HTML]{F3F3F3}-} &
  \multicolumn{1}{r|}{\cellcolor[HTML]{F3F3F3}0.00} &
  \multicolumn{1}{r|}{\cellcolor[HTML]{F3F3F3}0.52} &
  \multicolumn{1}{r|}{\cellcolor[HTML]{F3F3F3}0.00} &
  \multicolumn{1}{r|}{\cellcolor[HTML]{F3F3F3}0.00} &
  0.00 &
  \multicolumn{1}{r|}{\cellcolor[HTML]{F3F3F3}127} &
  \multicolumn{1}{r|}{\cellcolor[HTML]{F3F3F3}145} &
  \multicolumn{1}{r|}{\cellcolor[HTML]{F3F3F3}2} &
  \multicolumn{1}{r|}{\cellcolor[HTML]{F3F3F3}18} &
  \multicolumn{1}{r|}{\cellcolor[HTML]{D9EAD3}0.53} &
  \multicolumn{1}{r|}{\cellcolor[HTML]{D9EAD3}0.99} &
  \multicolumn{1}{r|}{\cellcolor[HTML]{D9EAD3}0.56} &
  \multicolumn{1}{r|}{\cellcolor[HTML]{D9EAD3}0.99} &
  \multicolumn{1}{r|}{\cellcolor[HTML]{D9EAD3}0.88} &
  \cellcolor[HTML]{D9EAD3}0.69 \\ \cline{2-25} 
\rowcolor[HTML]{F3F3F3} 
\cellcolor[HTML]{F3F3F3} &
  HP3 &
  45.45\% &
  \multicolumn{1}{r|}{\cellcolor[HTML]{F3F3F3}147} &
  145 &
  \multicolumn{1}{r|}{\cellcolor[HTML]{F3F3F3}0} &
  \multicolumn{1}{r|}{\cellcolor[HTML]{F3F3F3}0} &
  \multicolumn{1}{r|}{\cellcolor[HTML]{F3F3F3}139} &
  \multicolumn{1}{r|}{\cellcolor[HTML]{F3F3F3}153} &
  \multicolumn{1}{r|}{\cellcolor[HTML]{F3F3F3}-} &
  \multicolumn{1}{r|}{\cellcolor[HTML]{F3F3F3}0.00} &
  \multicolumn{1}{r|}{\cellcolor[HTML]{F3F3F3}0.52} &
  \multicolumn{1}{r|}{\cellcolor[HTML]{F3F3F3}0.00} &
  \multicolumn{1}{r|}{\cellcolor[HTML]{F3F3F3}0.00} &
  0.00 &
  \multicolumn{1}{r|}{\cellcolor[HTML]{F3F3F3}147} &
  \multicolumn{1}{r|}{\cellcolor[HTML]{F3F3F3}145} &
  \multicolumn{1}{r|}{\cellcolor[HTML]{F3F3F3}0} &
  \multicolumn{1}{r|}{\cellcolor[HTML]{F3F3F3}0} &
  \multicolumn{1}{r|}{\cellcolor[HTML]{F3F3F3}0.50} &
  \multicolumn{1}{r|}{\cellcolor[HTML]{F3F3F3}1.00} &
  \multicolumn{1}{r|}{\cellcolor[HTML]{F3F3F3}0.50} &
  \multicolumn{1}{r|}{\cellcolor[HTML]{F3F3F3}1.00} &
  \multicolumn{1}{r|}{\cellcolor[HTML]{F3F3F3}1.00} &
  0.66 \\ \cline{2-25} 
\rowcolor[HTML]{F3F3F3} 
\multirow{-4}{*}{\cellcolor[HTML]{F3F3F3}House Price} &
  HP4 &
  87.50\% &
  \multicolumn{1}{r|}{\cellcolor[HTML]{F3F3F3}145} &
  147 &
  \multicolumn{1}{r|}{\cellcolor[HTML]{F3F3F3}51} &
  \multicolumn{1}{r|}{\cellcolor[HTML]{F3F3F3}168} &
  \multicolumn{1}{r|}{\cellcolor[HTML]{F3F3F3}15} &
  \multicolumn{1}{r|}{\cellcolor[HTML]{F3F3F3}57} &
  \multicolumn{1}{r|}{\cellcolor[HTML]{F3F3F3}0.77} &
  \multicolumn{1}{r|}{\cellcolor[HTML]{F3F3F3}0.92} &
  \multicolumn{1}{r|}{\cellcolor[HTML]{F3F3F3}0.77} &
  \multicolumn{1}{r|}{\cellcolor[HTML]{F3F3F3}0.92} &
  \multicolumn{1}{r|}{\cellcolor[HTML]{F3F3F3}0.47} &
  0.84 &
  \multicolumn{1}{r|}{\cellcolor[HTML]{F3F3F3}143} &
  \multicolumn{1}{r|}{\cellcolor[HTML]{F3F3F3}146} &
  \multicolumn{1}{r|}{\cellcolor[HTML]{F3F3F3}1} &
  \multicolumn{1}{r|}{\cellcolor[HTML]{F3F3F3}2} &
  \multicolumn{1}{r|}{\cellcolor[HTML]{F3F3F3}0.51} &
  \multicolumn{1}{r|}{\cellcolor[HTML]{F3F3F3}0.99} &
  \multicolumn{1}{r|}{\cellcolor[HTML]{F3F3F3}0.51} &
  \multicolumn{1}{r|}{\cellcolor[HTML]{F3F3F3}0.99} &
  \multicolumn{1}{r|}{\cellcolor[HTML]{F3F3F3}0.99} &
  0.67 \\ \hline
 &
  BM1 &
  81.10\% &
  \multicolumn{1}{r|}{1044} &
  1072 &
  \multicolumn{1}{r|}{0} &
  \multicolumn{1}{r|}{0} &
  \multicolumn{1}{r|}{1024} &
  \multicolumn{1}{r|}{1092} &
  \multicolumn{1}{r|}{-} &
  \multicolumn{1}{r|}{0.00} &
  \multicolumn{1}{r|}{0.52} &
  \multicolumn{1}{r|}{0.00} &
  \multicolumn{1}{r|}{0.00} &
  0.00 &
  \multicolumn{1}{r|}{984} &
  \multicolumn{1}{r|}{987} &
  \multicolumn{1}{r|}{85} &
  \multicolumn{1}{r|}{60} &
  \multicolumn{1}{r|}{0.50} &
  \multicolumn{1}{r|}{0.92} &
  \multicolumn{1}{r|}{0.49} &
  \multicolumn{1}{r|}{0.92} &
  \multicolumn{1}{r|}{0.94} &
  0.65 \\ \cline{2-25} 
 &
  BM2 &
  82.11\% &
  \multicolumn{1}{r|}{1050} &
  1066 &
  \multicolumn{1}{r|}{0} &
  \multicolumn{1}{r|}{0} &
  \multicolumn{1}{r|}{1024} &
  \multicolumn{1}{r|}{1092} &
  \multicolumn{1}{r|}{-} &
  \multicolumn{1}{r|}{0.00} &
  \multicolumn{1}{r|}{0.52} &
  \multicolumn{1}{r|}{0.00} &
  \multicolumn{1}{r|}{0.00} &
  0.00 &
  \multicolumn{1}{r|}{869} &
  \multicolumn{1}{r|}{866} &
  \multicolumn{1}{r|}{200} &
  \multicolumn{1}{r|}{181} &
  \multicolumn{1}{r|}{0.50} &
  \multicolumn{1}{r|}{0.81} &
  \multicolumn{1}{r|}{0.49} &
  \multicolumn{1}{r|}{0.81} &
  \multicolumn{1}{r|}{0.83} &
  0.62 \\ \cline{2-25} 
 &
  BM3 &
  80.19\% &
  \multicolumn{1}{r|}{1024} &
  1092 &
  \multicolumn{1}{r|}{387} &
  \multicolumn{1}{r|}{798} &
  \multicolumn{1}{r|}{332} &
  \multicolumn{1}{r|}{599} &
  \multicolumn{1}{r|}{0.67} &
  \multicolumn{1}{r|}{0.71} &
  \multicolumn{1}{r|}{0.66} &
  \multicolumn{1}{r|}{0.71} &
  \multicolumn{1}{r|}{0.39} &
  0.69 &
  \multicolumn{1}{r|}{474} &
  \multicolumn{1}{r|}{559} &
  \multicolumn{1}{r|}{533} &
  \multicolumn{1}{r|}{550} &
  \multicolumn{1}{r|}{0.54} &
  \multicolumn{1}{r|}{0.51} &
  \multicolumn{1}{r|}{0.52} &
  \multicolumn{1}{r|}{0.51} &
  \multicolumn{1}{r|}{0.46} &
  0.53 \\ \cline{2-25} 
 &
  BM4 &
  79.10\% &
  \multicolumn{1}{r|}{1049} &
  1067 &
  \multicolumn{1}{r|}{0} &
  \multicolumn{1}{r|}{0} &
  \multicolumn{1}{r|}{1024} &
  \multicolumn{1}{r|}{1092} &
  \multicolumn{1}{r|}{-} &
  \multicolumn{1}{r|}{0.00} &
  \multicolumn{1}{r|}{0.52} &
  \multicolumn{1}{r|}{0.00} &
  \multicolumn{1}{r|}{0.00} &
  0.00 &
  \multicolumn{1}{r|}{916} &
  \multicolumn{1}{r|}{906} &
  \multicolumn{1}{r|}{161} &
  \multicolumn{1}{r|}{133} &
  \multicolumn{1}{r|}{0.50} &
  \multicolumn{1}{r|}{0.85} &
  \multicolumn{1}{r|}{0.49} &
  \multicolumn{1}{r|}{0.85} &
  \multicolumn{1}{r|}{0.87} &
  0.63 \\ \cline{2-25} 
 &
  BM5 &
  82.10\% &
  \multicolumn{1}{r|}{1064} &
  1052 &
  \multicolumn{1}{r|}{0} &
  \multicolumn{1}{r|}{0} &
  \multicolumn{1}{r|}{1024} &
  \multicolumn{1}{r|}{1092} &
  \multicolumn{1}{r|}{-} &
  \multicolumn{1}{r|}{0.00} &
  \multicolumn{1}{r|}{0.52} &
  \multicolumn{1}{r|}{0.00} &
  \multicolumn{1}{r|}{0.00} &
  0.00 &
  \multicolumn{1}{r|}{1001} &
  \multicolumn{1}{r|}{970} &
  \multicolumn{1}{r|}{82} &
  \multicolumn{1}{r|}{63} &
  \multicolumn{1}{r|}{0.49} &
  \multicolumn{1}{r|}{0.92} &
  \multicolumn{1}{r|}{0.49} &
  \multicolumn{1}{r|}{0.92} &
  \multicolumn{1}{r|}{0.94} &
  0.64 \\ \cline{2-25} 
 &
  BM6 &
  82.00\% &
  \multicolumn{1}{r|}{1057} &
  1059 &
  \multicolumn{1}{r|}{-} &
  \multicolumn{1}{r|}{-} &
  \multicolumn{1}{r|}{-} &
  \multicolumn{1}{r|}{-} &
  \multicolumn{1}{r|}{-} &
  \multicolumn{1}{r|}{-} &
  \multicolumn{1}{r|}{-} &
  \multicolumn{1}{r|}{-} &
  \multicolumn{1}{r|}{-} &
  - &
  \multicolumn{1}{r|}{1004} &
  \multicolumn{1}{r|}{977} &
  \multicolumn{1}{r|}{82} &
  \multicolumn{1}{r|}{53} &
  \multicolumn{1}{r|}{0.49} &
  \multicolumn{1}{r|}{0.92} &
  \multicolumn{1}{r|}{0.49} &
  \multicolumn{1}{r|}{0.92} &
  \multicolumn{1}{r|}{0.95} &
  0.64 \\ \cline{2-25} 
 &
  BM7 &
  81.00\% &
  \multicolumn{1}{r|}{1042} &
  1074 &
  \multicolumn{1}{r|}{-} &
  \multicolumn{1}{r|}{-} &
  \multicolumn{1}{r|}{-} &
  \multicolumn{1}{r|}{-} &
  \multicolumn{1}{r|}{-} &
  \multicolumn{1}{r|}{-} &
  \multicolumn{1}{r|}{-} &
  \multicolumn{1}{r|}{-} &
  \multicolumn{1}{r|}{-} &
  - &
  \multicolumn{1}{r|}{987} &
  \multicolumn{1}{r|}{984} &
  \multicolumn{1}{r|}{90} &
  \multicolumn{1}{r|}{55} &
  \multicolumn{1}{r|}{0.50} &
  \multicolumn{1}{r|}{0.92} &
  \multicolumn{1}{r|}{0.49} &
  \multicolumn{1}{r|}{0.92} &
  \multicolumn{1}{r|}{0.95} &
  0.65 \\ \cline{2-25} 
 &
  BM8 &
  82.00\% &
  \multicolumn{1}{r|}{1041} &
  1075 &
  \multicolumn{1}{r|}{0} &
  \multicolumn{1}{r|}{0} &
  \multicolumn{1}{r|}{1024} &
  \multicolumn{1}{r|}{1092} &
  \multicolumn{1}{r|}{-} &
  \multicolumn{1}{r|}{-} &
  \multicolumn{1}{r|}{-} &
  \multicolumn{1}{r|}{-} &
  \multicolumn{1}{r|}{-} &
  - &
  \multicolumn{1}{r|}{985} &
  \multicolumn{1}{r|}{986} &
  \multicolumn{1}{r|}{89} &
  \multicolumn{1}{r|}{56} &
  \multicolumn{1}{r|}{0.50} &
  \multicolumn{1}{r|}{0.92} &
  \multicolumn{1}{r|}{0.49} &
  \multicolumn{1}{r|}{0.92} &
  \multicolumn{1}{r|}{0.95} &
  0.65 \\ \cline{2-25} 
 &
  BM9 &
  81.30\% &
  \multicolumn{1}{r|}{1058} &
  1058 &
  \multicolumn{1}{r|}{-} &
  \multicolumn{1}{r|}{-} &
  \multicolumn{1}{r|}{-} &
  \multicolumn{1}{r|}{-} &
  \multicolumn{1}{r|}{-} &
  \multicolumn{1}{r|}{-} &
  \multicolumn{1}{r|}{-} &
  \multicolumn{1}{r|}{-} &
  \multicolumn{1}{r|}{-} &
  - &
  \multicolumn{1}{r|}{923} &
  \multicolumn{1}{r|}{888} &
  \multicolumn{1}{r|}{170} &
  \multicolumn{1}{r|}{135} &
  \multicolumn{1}{r|}{0.49} &
  \multicolumn{1}{r|}{0.84} &
  \multicolumn{1}{r|}{0.48} &
  \multicolumn{1}{r|}{0.84} &
  \multicolumn{1}{r|}{0.87} &
  0.62 \\ \cline{2-25} 
 &
  BM10 &
  81.90\% &
  \multicolumn{1}{r|}{1024} &
  1092 &
  \multicolumn{1}{r|}{-} &
  \multicolumn{1}{r|}{-} &
  \multicolumn{1}{r|}{-} &
  \multicolumn{1}{r|}{-} &
  \multicolumn{1}{r|}{-} &
  \multicolumn{1}{r|}{-} &
  \multicolumn{1}{r|}{-} &
  \multicolumn{1}{r|}{-} &
  \multicolumn{1}{r|}{-} &
  - &
  \multicolumn{1}{r|}{982} &
  \multicolumn{1}{r|}{989} &
  \multicolumn{1}{r|}{103} &
  \multicolumn{1}{r|}{42} &
  \multicolumn{1}{r|}{0.50} &
  \multicolumn{1}{r|}{0.91} &
  \multicolumn{1}{r|}{0.49} &
  \multicolumn{1}{r|}{0.91} &
  \multicolumn{1}{r|}{0.96} &
  0.65 \\ \cline{2-25} 
 &
  \cellcolor[HTML]{D9EAD3}BM11 &
  \cellcolor[HTML]{D9EAD3}83.60\% &
  \multicolumn{1}{r|}{1062} &
  1054 &
  \multicolumn{1}{r|}{-} &
  \multicolumn{1}{r|}{-} &
  \multicolumn{1}{r|}{-} &
  \multicolumn{1}{r|}{-} &
  \multicolumn{1}{r|}{-} &
  \multicolumn{1}{r|}{-} &
  \multicolumn{1}{r|}{-} &
  \multicolumn{1}{r|}{-} &
  \multicolumn{1}{r|}{-} &
  - &
  \multicolumn{1}{r|}{1062} &
  \multicolumn{1}{r|}{1054} &
  \multicolumn{1}{r|}{0} &
  \multicolumn{1}{r|}{0} &
  \multicolumn{1}{r|}{\cellcolor[HTML]{D9EAD3}0.50} &
  \multicolumn{1}{r|}{\cellcolor[HTML]{D9EAD3}1.00} &
  \multicolumn{1}{r|}{\cellcolor[HTML]{D9EAD3}0.50} &
  \multicolumn{1}{r|}{\cellcolor[HTML]{D9EAD3}1.00} &
  \multicolumn{1}{r|}{\cellcolor[HTML]{D9EAD3}1.00} &
  \cellcolor[HTML]{D9EAD3}0.66 \\
  \cline{2-25} 
\multirow{-12}{*}{BankCustomer} &
  BM12 &
  80.70\% &
  \multicolumn{1}{r|}{1041} &
  1075 &
  \multicolumn{1}{r|}{0} &
  \multicolumn{1}{r|}{0} &
  \multicolumn{1}{r|}{1024} &
  \multicolumn{1}{r|}{1092} &
  \multicolumn{1}{r|}{-} &
  \multicolumn{1}{r|}{0.00} &
  \multicolumn{1}{r|}{0.52} &
  \multicolumn{1}{r|}{0.00} &
  \multicolumn{1}{r|}{0.00} &
  0.00 &
  \multicolumn{1}{r|}{860} &
  \multicolumn{1}{r|}{866} &
  \multicolumn{1}{r|}{209} &
  \multicolumn{1}{r|}{181} &
  \multicolumn{1}{r|}{0.50} &
  \multicolumn{1}{r|}{0.81} &
  \multicolumn{1}{r|}{0.49} &
  \multicolumn{1}{r|}{0.81} &
  \multicolumn{1}{r|}{0.83} &
  0.62 \\ \hline
\rowcolor[HTML]{F3F3F3} 
\cellcolor[HTML]{F3F3F3} &
  \cellcolor[HTML]{D9EAD3}GC1 &
  \cellcolor[HTML]{D9EAD3}99.00\% &
  \multicolumn{1}{r|}{\cellcolor[HTML]{F3F3F3}2} &
  198 &
  \multicolumn{1}{r|}{\cellcolor[HTML]{F3F3F3}37} &
  \multicolumn{1}{r|}{\cellcolor[HTML]{F3F3F3}91} &
  \multicolumn{1}{r|}{\cellcolor[HTML]{F3F3F3}1} &
  \multicolumn{1}{r|}{\cellcolor[HTML]{F3F3F3}71} &
  \multicolumn{1}{r|}{\cellcolor[HTML]{F3F3F3}0.71} &
  \multicolumn{1}{r|}{\cellcolor[HTML]{F3F3F3}0.99} &
  \multicolumn{1}{r|}{\cellcolor[HTML]{F3F3F3}0.81} &
  \multicolumn{1}{r|}{\cellcolor[HTML]{F3F3F3}0.99} &
  \multicolumn{1}{r|}{\cellcolor[HTML]{F3F3F3}0.34} &
  0.83 &
  \multicolumn{1}{r|}{\cellcolor[HTML]{F3F3F3}2} &
  \multicolumn{1}{r|}{\cellcolor[HTML]{F3F3F3}198} &
  \multicolumn{1}{r|}{\cellcolor[HTML]{F3F3F3}0} &
  \multicolumn{1}{r|}{\cellcolor[HTML]{F3F3F3}0} &
  \multicolumn{1}{r|}{\cellcolor[HTML]{D9EAD3}0.99} &
  \multicolumn{1}{r|}{\cellcolor[HTML]{D9EAD3}1.00} &
  \multicolumn{1}{r|}{\cellcolor[HTML]{D9EAD3}0.99} &
  \multicolumn{1}{r|}{\cellcolor[HTML]{D9EAD3}1.00} &
  \multicolumn{1}{r|}{\cellcolor[HTML]{D9EAD3}1.00} &
  \cellcolor[HTML]{D9EAD3}0.99 \\ \cline{2-25} 
\rowcolor[HTML]{F3F3F3} 
\cellcolor[HTML]{F3F3F3} &
  GC2 &
  99.00\% &
  \multicolumn{1}{r|}{\cellcolor[HTML]{F3F3F3}2} &
  198 &
  \multicolumn{1}{r|}{\cellcolor[HTML]{F3F3F3}0} &
  \multicolumn{1}{r|}{\cellcolor[HTML]{F3F3F3}74} &
  \multicolumn{1}{r|}{\cellcolor[HTML]{F3F3F3}18} &
  \multicolumn{1}{r|}{\cellcolor[HTML]{F3F3F3}108} &
  \multicolumn{1}{r|}{\cellcolor[HTML]{F3F3F3}1.00} &
  \multicolumn{1}{r|}{\cellcolor[HTML]{F3F3F3}0.80} &
  \multicolumn{1}{r|}{\cellcolor[HTML]{F3F3F3}0.91} &
  \multicolumn{1}{r|}{\cellcolor[HTML]{F3F3F3}0.80} &
  \multicolumn{1}{r|}{\cellcolor[HTML]{F3F3F3}0.00} &
  0.89 &
  \multicolumn{1}{r|}{\cellcolor[HTML]{F3F3F3}2} &
  \multicolumn{1}{r|}{\cellcolor[HTML]{F3F3F3}186} &
  \multicolumn{1}{r|}{\cellcolor[HTML]{F3F3F3}12} &
  \multicolumn{1}{r|}{\cellcolor[HTML]{F3F3F3}0} &
  \multicolumn{1}{r|}{\cellcolor[HTML]{F3F3F3}0.99} &
  \multicolumn{1}{r|}{\cellcolor[HTML]{F3F3F3}0.94} &
  \multicolumn{1}{r|}{\cellcolor[HTML]{F3F3F3}0.93} &
  \multicolumn{1}{r|}{\cellcolor[HTML]{F3F3F3}0.94} &
  \multicolumn{1}{r|}{\cellcolor[HTML]{F3F3F3}1.00} &
  0.96 \\ \cline{2-25} 
\rowcolor[HTML]{F3F3F3} 
\cellcolor[HTML]{F3F3F3} &
  GC3 &
  99.00\% &
  \multicolumn{1}{r|}{\cellcolor[HTML]{F3F3F3}2} &
  198 &
  \multicolumn{1}{r|}{\cellcolor[HTML]{F3F3F3}42} &
  \multicolumn{1}{r|}{\cellcolor[HTML]{F3F3F3}96} &
  \multicolumn{1}{r|}{\cellcolor[HTML]{F3F3F3}1} &
  \multicolumn{1}{r|}{\cellcolor[HTML]{F3F3F3}61} &
  \multicolumn{1}{r|}{\cellcolor[HTML]{F3F3F3}0.70} &
  \multicolumn{1}{r|}{\cellcolor[HTML]{F3F3F3}0.99} &
  \multicolumn{1}{r|}{\cellcolor[HTML]{F3F3F3}0.79} &
  \multicolumn{1}{r|}{\cellcolor[HTML]{F3F3F3}0.99} &
  \multicolumn{1}{r|}{\cellcolor[HTML]{F3F3F3}0.41} &
  0.82 &
  \multicolumn{1}{r|}{\cellcolor[HTML]{F3F3F3}1} &
  \multicolumn{1}{r|}{\cellcolor[HTML]{F3F3F3}72} &
  \multicolumn{1}{r|}{\cellcolor[HTML]{F3F3F3}126} &
  \multicolumn{1}{r|}{\cellcolor[HTML]{F3F3F3}1} &
  \multicolumn{1}{r|}{\cellcolor[HTML]{F3F3F3}0.99} &
  \multicolumn{1}{r|}{\cellcolor[HTML]{F3F3F3}0.36} &
  \multicolumn{1}{r|}{\cellcolor[HTML]{F3F3F3}0.37} &
  \multicolumn{1}{r|}{\cellcolor[HTML]{F3F3F3}0.36} &
  \multicolumn{1}{r|}{\cellcolor[HTML]{F3F3F3}0.50} &
  0.53 \\ \cline{2-25} 
\rowcolor[HTML]{F3F3F3} 
\cellcolor[HTML]{F3F3F3} &
  \cellcolor[HTML]{D9EAD3}GC4 &
  \cellcolor[HTML]{D9EAD3}99.00\% &
  \multicolumn{1}{r|}{\cellcolor[HTML]{F3F3F3}2} &
  198 &
  \multicolumn{1}{r|}{\cellcolor[HTML]{F3F3F3}0} &
  \multicolumn{1}{r|}{\cellcolor[HTML]{F3F3F3}2} &
  \multicolumn{1}{r|}{\cellcolor[HTML]{F3F3F3}0} &
  \multicolumn{1}{r|}{\cellcolor[HTML]{F3F3F3}198} &
  \multicolumn{1}{r|}{\cellcolor[HTML]{F3F3F3}1.00} &
  \multicolumn{1}{r|}{\cellcolor[HTML]{F3F3F3}1.00} &
  \multicolumn{1}{r|}{\cellcolor[HTML]{F3F3F3}1.00} &
  \multicolumn{1}{r|}{\cellcolor[HTML]{F3F3F3}1.00} &
  \multicolumn{1}{r|}{\cellcolor[HTML]{F3F3F3}0.00} &
  1.00 &
  \multicolumn{1}{r|}{\cellcolor[HTML]{F3F3F3}2} &
  \multicolumn{1}{r|}{\cellcolor[HTML]{F3F3F3}198} &
  \multicolumn{1}{r|}{\cellcolor[HTML]{F3F3F3}0} &
  \multicolumn{1}{r|}{\cellcolor[HTML]{F3F3F3}0} &
  \multicolumn{1}{r|}{\cellcolor[HTML]{D9EAD3}0.99} &
  \multicolumn{1}{r|}{\cellcolor[HTML]{D9EAD3}1.00} &
  \multicolumn{1}{r|}{\cellcolor[HTML]{D9EAD3}0.99} &
  \multicolumn{1}{r|}{\cellcolor[HTML]{D9EAD3}1.00} &
  \multicolumn{1}{r|}{\cellcolor[HTML]{D9EAD3}1.00} &
  \cellcolor[HTML]{D9EAD3}0.99 \\ \cline{2-25} 
\rowcolor[HTML]{F3F3F3} 
\cellcolor[HTML]{F3F3F3} &
  GC5 &
  99.00\% &
  \multicolumn{1}{r|}{\cellcolor[HTML]{F3F3F3}2} &
  198 &
  \multicolumn{1}{r|}{\cellcolor[HTML]{F3F3F3}0} &
  \multicolumn{1}{r|}{\cellcolor[HTML]{F3F3F3}0} &
  \multicolumn{1}{r|}{\cellcolor[HTML]{F3F3F3}2} &
  \multicolumn{1}{r|}{\cellcolor[HTML]{F3F3F3}198} &
  \multicolumn{1}{r|}{\cellcolor[HTML]{F3F3F3}-} &
  \multicolumn{1}{r|}{\cellcolor[HTML]{F3F3F3}0.00} &
  \multicolumn{1}{r|}{\cellcolor[HTML]{F3F3F3}0.99} &
  \multicolumn{1}{r|}{\cellcolor[HTML]{F3F3F3}0.00} &
  \multicolumn{1}{r|}{\cellcolor[HTML]{F3F3F3}0.00} &
  0.00 &
  \multicolumn{1}{r|}{\cellcolor[HTML]{F3F3F3}2} &
  \multicolumn{1}{r|}{\cellcolor[HTML]{F3F3F3}193} &
  \multicolumn{1}{r|}{\cellcolor[HTML]{F3F3F3}5} &
  \multicolumn{1}{r|}{\cellcolor[HTML]{F3F3F3}0} &
  \multicolumn{1}{r|}{\cellcolor[HTML]{F3F3F3}0.99} &
  \multicolumn{1}{r|}{\cellcolor[HTML]{F3F3F3}0.97} &
  \multicolumn{1}{r|}{\cellcolor[HTML]{F3F3F3}0.97} &
  \multicolumn{1}{r|}{\cellcolor[HTML]{F3F3F3}0.97} &
  \multicolumn{1}{r|}{\cellcolor[HTML]{F3F3F3}1.00} &
  0.98 \\ \cline{2-25} 
\rowcolor[HTML]{F3F3F3} 
\cellcolor[HTML]{F3F3F3} &
  GC6 &
  99.00\% &
  \multicolumn{1}{r|}{\cellcolor[HTML]{F3F3F3}2} &
  198 &
  \multicolumn{1}{r|}{\cellcolor[HTML]{F3F3F3}0} &
  \multicolumn{1}{r|}{\cellcolor[HTML]{F3F3F3}0} &
  \multicolumn{1}{r|}{\cellcolor[HTML]{F3F3F3}2} &
  \multicolumn{1}{r|}{\cellcolor[HTML]{F3F3F3}198} &
  \multicolumn{1}{r|}{\cellcolor[HTML]{F3F3F3}-} &
  \multicolumn{1}{r|}{\cellcolor[HTML]{F3F3F3}0.00} &
  \multicolumn{1}{r|}{\cellcolor[HTML]{F3F3F3}0.99} &
  \multicolumn{1}{r|}{\cellcolor[HTML]{F3F3F3}0.00} &
  \multicolumn{1}{r|}{\cellcolor[HTML]{F3F3F3}0.00} &
  0.00 &
  \multicolumn{1}{r|}{\cellcolor[HTML]{F3F3F3}1} &
  \multicolumn{1}{r|}{\cellcolor[HTML]{F3F3F3}66} &
  \multicolumn{1}{r|}{\cellcolor[HTML]{F3F3F3}132} &
  \multicolumn{1}{r|}{\cellcolor[HTML]{F3F3F3}1} &
  \multicolumn{1}{r|}{\cellcolor[HTML]{F3F3F3}0.99} &
  \multicolumn{1}{r|}{\cellcolor[HTML]{F3F3F3}0.33} &
  \multicolumn{1}{r|}{\cellcolor[HTML]{F3F3F3}0.34} &
  \multicolumn{1}{r|}{\cellcolor[HTML]{F3F3F3}0.33} &
  \multicolumn{1}{r|}{\cellcolor[HTML]{F3F3F3}0.50} &
  0.50 \\ \cline{2-25} 
\rowcolor[HTML]{F3F3F3} 
\cellcolor[HTML]{F3F3F3} &
  GC7 &
  99.00\% &
  \multicolumn{1}{r|}{\cellcolor[HTML]{F3F3F3}2} &
  198 &
  \multicolumn{1}{r|}{\cellcolor[HTML]{F3F3F3}0} &
  \multicolumn{1}{r|}{\cellcolor[HTML]{F3F3F3}0} &
  \multicolumn{1}{r|}{\cellcolor[HTML]{F3F3F3}2} &
  \multicolumn{1}{r|}{\cellcolor[HTML]{F3F3F3}198} &
  \multicolumn{1}{r|}{\cellcolor[HTML]{F3F3F3}-} &
  \multicolumn{1}{r|}{\cellcolor[HTML]{F3F3F3}0.00} &
  \multicolumn{1}{r|}{\cellcolor[HTML]{F3F3F3}0.99} &
  \multicolumn{1}{r|}{\cellcolor[HTML]{F3F3F3}0.00} &
  \multicolumn{1}{r|}{\cellcolor[HTML]{F3F3F3}0.00} &
  0.00 &
  \multicolumn{1}{r|}{\cellcolor[HTML]{F3F3F3}2} &
  \multicolumn{1}{r|}{\cellcolor[HTML]{F3F3F3}193} &
  \multicolumn{1}{r|}{\cellcolor[HTML]{F3F3F3}5} &
  \multicolumn{1}{r|}{\cellcolor[HTML]{F3F3F3}0} &
  \multicolumn{1}{r|}{\cellcolor[HTML]{F3F3F3}0.99} &
  \multicolumn{1}{r|}{\cellcolor[HTML]{F3F3F3}0.97} &
  \multicolumn{1}{r|}{\cellcolor[HTML]{F3F3F3}0.97} &
  \multicolumn{1}{r|}{\cellcolor[HTML]{F3F3F3}0.97} &
  \multicolumn{1}{r|}{\cellcolor[HTML]{F3F3F3}1.00} &
  0.98 \\ \cline{2-25} 
\rowcolor[HTML]{F3F3F3} 
\cellcolor[HTML]{F3F3F3} &
  GC8 &
  99.00\% &
  \multicolumn{1}{r|}{\cellcolor[HTML]{F3F3F3}2} &
  198 &
  \multicolumn{1}{r|}{\cellcolor[HTML]{F3F3F3}0} &
  \multicolumn{1}{r|}{\cellcolor[HTML]{F3F3F3}0} &
  \multicolumn{1}{r|}{\cellcolor[HTML]{F3F3F3}2} &
  \multicolumn{1}{r|}{\cellcolor[HTML]{F3F3F3}198} &
  \multicolumn{1}{r|}{\cellcolor[HTML]{F3F3F3}-} &
  \multicolumn{1}{r|}{\cellcolor[HTML]{F3F3F3}0.00} &
  \multicolumn{1}{r|}{\cellcolor[HTML]{F3F3F3}0.99} &
  \multicolumn{1}{r|}{\cellcolor[HTML]{F3F3F3}0.00} &
  \multicolumn{1}{r|}{\cellcolor[HTML]{F3F3F3}0.00} &
  0.00 &
  \multicolumn{1}{r|}{\cellcolor[HTML]{F3F3F3}1} &
  \multicolumn{1}{r|}{\cellcolor[HTML]{F3F3F3}142} &
  \multicolumn{1}{r|}{\cellcolor[HTML]{F3F3F3}56} &
  \multicolumn{1}{r|}{\cellcolor[HTML]{F3F3F3}1} &
  \multicolumn{1}{r|}{\cellcolor[HTML]{F3F3F3}0.99} &
  \multicolumn{1}{r|}{\cellcolor[HTML]{F3F3F3}0.72} &
  \multicolumn{1}{r|}{\cellcolor[HTML]{F3F3F3}0.72} &
  \multicolumn{1}{r|}{\cellcolor[HTML]{F3F3F3}0.72} &
  \multicolumn{1}{r|}{\cellcolor[HTML]{F3F3F3}0.50} &
  0.83 \\ \cline{2-25} 
\rowcolor[HTML]{F3F3F3} 
\multirow{-9}{*}{\cellcolor[HTML]{F3F3F3}GermanCredit} &
  GC9 &
  99.00\% &
  \multicolumn{1}{r|}{\cellcolor[HTML]{F3F3F3}2} &
  198 &
  \multicolumn{1}{r|}{\cellcolor[HTML]{F3F3F3}0} &
  \multicolumn{1}{r|}{\cellcolor[HTML]{F3F3F3}0} &
  \multicolumn{1}{r|}{\cellcolor[HTML]{F3F3F3}2} &
  \multicolumn{1}{r|}{\cellcolor[HTML]{F3F3F3}198} &
  \multicolumn{1}{r|}{\cellcolor[HTML]{F3F3F3}-} &
  \multicolumn{1}{r|}{\cellcolor[HTML]{F3F3F3}0.00} &
  \multicolumn{1}{r|}{\cellcolor[HTML]{F3F3F3}0.99} &
  \multicolumn{1}{r|}{\cellcolor[HTML]{F3F3F3}0.00} &
  \multicolumn{1}{r|}{\cellcolor[HTML]{F3F3F3}0.00} &
  0.00 &
  \multicolumn{1}{r|}{\cellcolor[HTML]{F3F3F3}1} &
  \multicolumn{1}{r|}{\cellcolor[HTML]{F3F3F3}142} &
  \multicolumn{1}{r|}{\cellcolor[HTML]{F3F3F3}56} &
  \multicolumn{1}{r|}{\cellcolor[HTML]{F3F3F3}1} &
  \multicolumn{1}{r|}{\cellcolor[HTML]{F3F3F3}0.99} &
  \multicolumn{1}{r|}{\cellcolor[HTML]{F3F3F3}0.72} &
  \multicolumn{1}{r|}{\cellcolor[HTML]{F3F3F3}0.72} &
  \multicolumn{1}{r|}{\cellcolor[HTML]{F3F3F3}0.72} &
  \multicolumn{1}{r|}{\cellcolor[HTML]{F3F3F3}0.50} &
  0.83 \\ \hline
\end{tabular}%
\label{tab:rq2}
}
 \footnotesize
  { * Here, `-' in "FP", "TP", "FN", "TN" column indicates where \selfchecker does not provide any output, therefore we can not get any values. For those cases, we get divided by zero error in the "Precision", "Recall", "Accuracy", "TPR", "FPR", "F-1" columns.}
\end{table*}

\begin{figure}[!h]
\centering
	\includegraphics[width=3.2in,trim={1.2cm 1.2cm 0cm 1.2cm},clip]{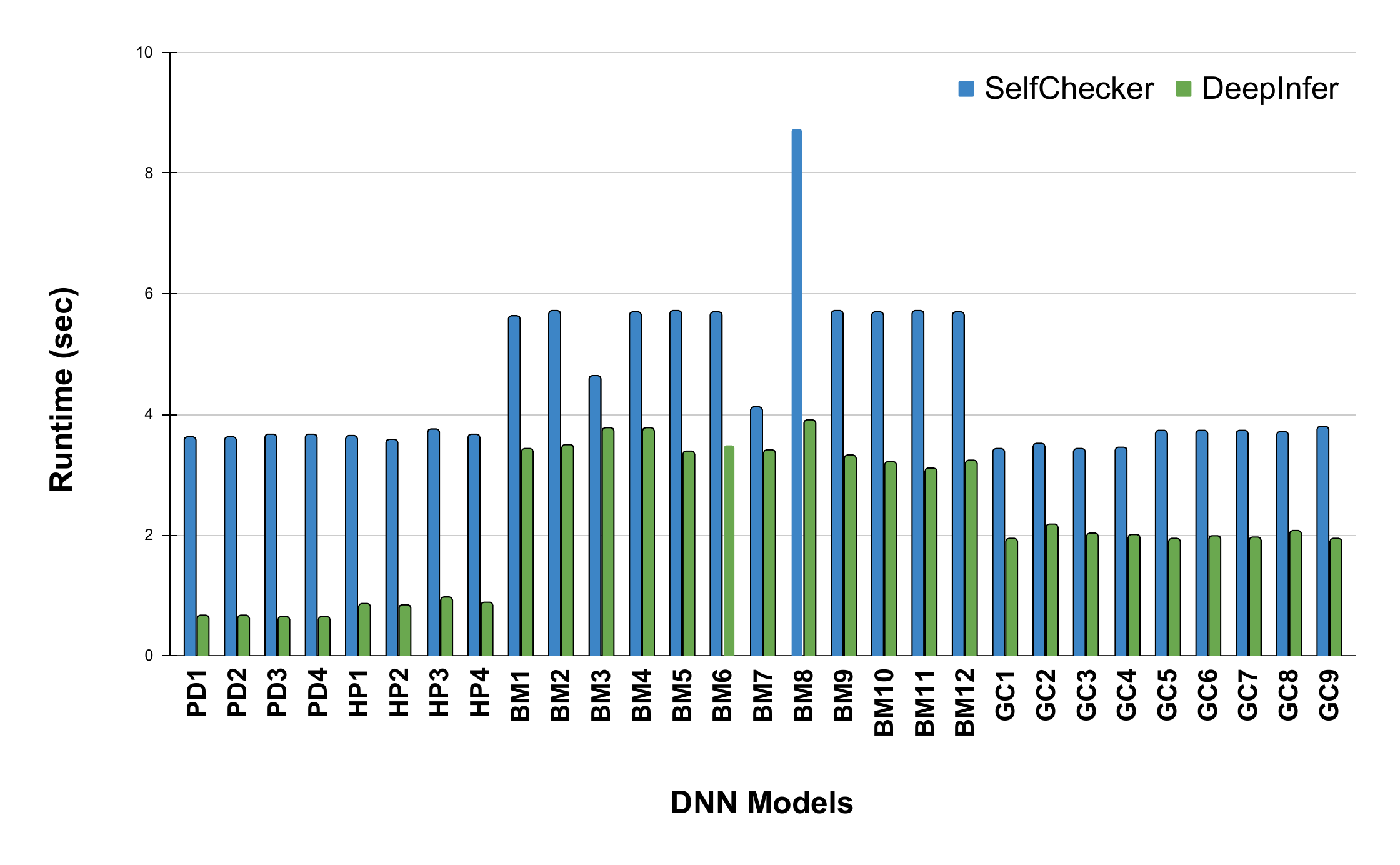}
	\footnotesize
	\caption{Runtime comparison of \deepinfer and \selfchecker for all models across different datasets} 
	\label{fig:rq3first}
\end{figure}
 
Next, we compute the precision, recall, and accuracy for all the models and present the results in Table~\ref{tab:rq2}. We computed average precision, recall, and accuracy for each dataset and obtained that for the high-accuracy models, the average precision, recall, and accuracy are  0.76, 0.98, and 0.76, respectively. Higher precision means that \deepinfer implies accurate results than inaccurate ones, and high recall means that \deepinfer returns most of the accurate results. 
The average precision and accuracy are low for models with 
performance-related underlying issues, which calls for further research. 
Furthermore, we also compared against the \selfchecker and found that \selfchecker produced identical results in terms of TP, FP, FN, and TN for certain models on a specific dataset. However, the assumption of using density functions and selected layers in the training module might not work properly. Also, measuring density function using training and representative test datasets might not be independent of model architectures, and it might not work well on different model structures which learned the training data differently.

\textit{In summary, \deepinfer effectively implies the correct and incorrect prediction of higher accuracy models with recall (0.98) and F-1 score (0.84), compared to \selfchecker with recall (0.59) and F-1 score (0.52).}
\begin{figure}[!h]
\centering
	\includegraphics[width=3in,trim={0.5cm 0.5cm 1cm 0.5cm},clip]{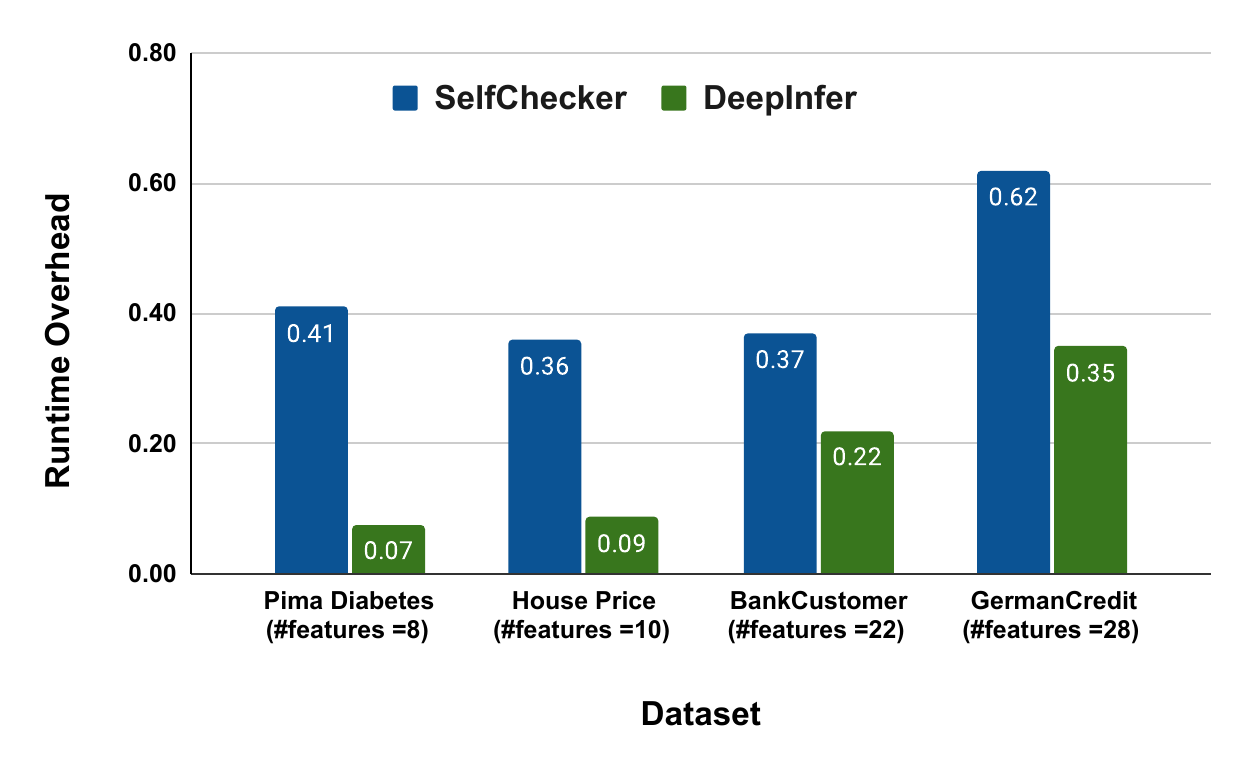}
	\footnotesize
	\caption{Runtime overhead comparison of \deepinfer and \selfchecker for all unseen data} 
	\label{fig:rq3second}
\end{figure}

\textbf{RQ3 (Efficiency):}  
We computed the runtime overhead of \deepinfer and \selfchecker with respect to original training time for all models in each kind of dataset which are unseen and plotted in Fig.~\ref{fig:rq3first}.  
From the results, we observed that the average runtime of \deepinfer is 0.66 sec, 0.88 sec, 3.46 sec, 2.00 sec compared to average training time of 8.88 sec, 10.15 sec, 15.67 sec, and 5.74 sec in Pima Diabetes, House Price, Bank Customer, German Credit dataset respectively. 
On the other hand, the average runtime of \selfchecker is 3.65, 3.66, 5.73, and 3.61 sec using all the models of Pima Diabetes, House Price, Bank Customer, and German Credit dataset, respectively.
We observe that the runtime is proportional to the number of features, which is 
consistent with our theoretical complexity results. 
Furthermore, we computed the runtime overhead of \deepinfer and \selfchecker 
for all unseen datasets over the training time for all models in each kind 
of dataset and plotted it in Fig.~\ref{fig:rq3second}.
We have observed that, the average runtime overhead of \selfchecker and \deepinfer is 0.41 and 0.07, 0.36 and 0.09, 0.37 and 0.22, 0.62 and 0.35 respectively, for Pima Diabetes, House Price, Bank Customer, German Credit dataset. 
During the deployment phase, we found that \deepinfer outperforms \selfchecker in terms of speed, being approximately 3.27 times faster. Additionally, we calculated the average runtime overhead for all unseen datasets and models, which is 0.22 seconds. This runtime overhead is relatively minimal when compared to the original training time. An advantage of our proposed approach is that we eliminate the need to repeatedly retrain the model for overhead computation. In contrast, \selfchecker requires extensive computations for all training and test datasets, along with different layer combinations, in order to calculate statistical measures like KDE values. Consequently, this process incurs a substantial runtime overhead. 

\textit{In summary, the average runtime overhead of \deepinfer is fairly minimal (0.22 sec for all the unseen data). The runtime overhead of \deepinfer is 3.27 times faster than \selfchecker during deployment.}
\subsubsection{Limitation}  
In this study, we conducted experiments to evaluate our proposed technique for inferring preconditions from real-valued features. We focused on these features because they are easier for humans to understand, and our datasets only included numerical values. While our current algorithms and derived \wpre rules are specific to certain layer computations and activation functions of fully connected layers, we believe that the fundamental idea of inferring data preconditions from deep neural network (DNN) models after training and using them for trustworthy prediction in deployment can be applied to other types of DNNs. For example, in popular models that utilize convolution and attention layers, we can extend the concept of computing data preconditions by extracting features from raw input data, such as images or text, and inferring preconditions from the classifier similarly.

\subsubsection{Discussion on the state-of-the-art (SOTA) metrics and approaches}
Some classifiers produce a confidence measure, such as confidence score and class prediction, typically by applying a softmax function to the raw numeric prediction values. However, such confidence measures need to be better-calibrated~\cite{jiang2018trust}; therefore, they cannot be reliably used as a measure of trust in prediction~\cite{fariha2021conformance}. 
Surprise coverage relies on the concept of surprise adequacy~\cite{weiss2022simple, surpriseTosem}, which measures the dissimilarity between a test and the training data set. Surprise adequacy has a high computational cost.
Surprise adequacy aims to quantitatively measure how surprising each new test input is when compared to the training data. It is used to detect out-of-bound with respect to the distribution of the training data, and the input is also more likely to cause unexpected model behavior. However, given an input, it captures the activation trace, the collection of neuron outputs produced by the model under test, which is expensive even for a simple model. Moreover, it does not indicate whether a particular prediction of the model is correct or incorrect with an unseen data point.
DeepGini score~\cite{feng2020deepgini} mainly provides a way to calculate a test prioritization to improve the quality of DNN. It determines a score by using only on the test input activations of the DNN's softmax output layer, limiting the approach’s applicability to only classification problems with softmax activation function in the last layer. Moreover, it does not provide a mechanism to imply whether a particular prediction of the model is correct or incorrect during deployment.
Some classifiers provide a level of confidence~\cite{confidenceRef} or certainty when making predictions about which class something belongs to. They usually calculate this confidence using the softmax function. However, these confidence scores are often not very accurate and can't be trusted to tell us how confident the classifier is about its prediction and imply whether it is correct or incorrect prediction.
None of these SOTA metrics learns input constraints from the trained model and utilizes that during the deployment to imply trust in the model's prediction using unseen data.
For the evaluation with publicly available fully connected DNNs and datasets with numerical values, the SOTA techniques \textit{SELFORACLE}~\cite{selforacle}, \textit{DISSECTOR}~\cite{dissector}, \textit{ConfidNet}~\cite{confidnet} are not applicable (details in~\secref{sec:related}).
\section{Related Work}
\label{sec:related}

We are inspired by the vast body of seminal work on weakest precondition calculus~\cite{dijkstra1975guarded, hoare1969axiomatic, hoare1987weakest, poskitt2010hoare, bonsangue1994weakest, de1999wp, d2006quantum, ying2012floyd}. 

\textbf{Trusted Machine Learning.}
The closest idea related to trusted machine learning in the database and machine learning community is Conformance Constraint Discovery (\ccsynth)~\cite{fariha2021conformance} to quantify the degree of non-conformance in a dataset, allowing for the effective characterization of whether or not inference over a given tuple is reliable. 
They demonstrated the application for detecting unsafe tuples in trustworthy machine learning. However, their approach is model-independent and will result in the same constraints for different models with the same dataset. Our approach resolves this issue and works as a model-specific approach to identify how to imply trust in different DL models' predictions using a dataset with unseen data during deployment. 
In the software engineering community, 
\textit{SELFORACLE}~\cite{selforacle} has proposed an approach that monitors the performance 
of the DNN at runtime to predict unsupported driving scenarios by computing a confidence estimation.
In contrast, our approach produces preconditions from the model using offline computation.  
\textit{SELFORACLE} also focuses on image-based models and temporally ordered inputs, 
such as video frames, and does not apply to data with numerical attributes.
Another technique, \selfchecker~\cite{selfchecker}, assesses model consistency during deployment and assumes that the density functions and layers chosen by the training module can be applicable to new test instances. However, this assumption is contingent upon whether the training and validation datasets accurately represent the characteristics of test instances. \selfchecker operates through a layer-based approach, which necessitates white-box access and may have limited capabilities in detecting issues in shallow DNNs with a few layers. \textit{SelfChecker++}~\cite{selfcheckerPlus} has been designed to target both unintended abnormal test data and intended adversarial samples. \textit{InputReflector}~\cite{InputReflector}, introduced a runtime approach to identify and fix failure-inducing inputs in DL systems inspired by traditional input-debugging techniques.
Wang et al. introduced \textit{DISSECTOR}~\cite{dissector}
to identify inputs that deviate from the norm, by training several sub-models on top of a pre-trained deep learning model. However, generating these sub-models is manual and time-consuming~\cite{selfchecker}.
Further, \textit{DISSECTOR} is only applicable to image-based models such as ImageNet~\cite{Dissectorrepo}. 
Researchers in the deep learning community have developed learning-based models to measure a model's confidence during deployment~\cite{confidnet, dlconf1, dlconf2, dlconf3, jiang2018trust, luo2021learning}. However, these models can be untrustworthy and suffer from overfitting.
Corbière et al.~\cite{confidnet} proposed \textit{ConfidNet}, a model built on 
top of pre-trained models that uses true class probability for failure prediction.  
However, overfitting can occur due to being trained on a small number of incorrect 
predictions in training dataset. 
\textit{ConfidNet} technique has ConvNet architecture in its implementation 
and it would not be applicable for DNNs with only dense layers and datasets with numerical values.
In contrast, our approach infers the model's assumption of the data after training and utilizes that to imply the trustworthiness of model's prediction.

\textbf{Neural Network Abstraction.}
There are a number of research ideas that focuses on abstracting neural network as DNN verification is NP-hard due to the number of nodes in DNN slowing the algorithms exponentially~\cite{abstract}.
Singh~\etal~\cite{abstractdomain} proposes an abstract domain based on floating-point polyhedra and intervals along with abstract transformers for neural network functions for certifying deep neural networks. Gehr~\etal~\cite{ai2} introduces the idea of abstract transformers that capture the behavior of common neural network layers to certify convolutional and large fully connected networks. 
There are other abstractions of neural networks, e.g., interval universal approximation~\cite{wang2022interval}, neural interval abstraction, neural zonotope abstraction, and neural polyhedron abstraction~\cite{aws}
None of these abstractions of the neural network works for \wpre reasoning with neural network functions as code statements and expected output as a postcondition which \deepinfer demonstrates.

\textbf{Neural Network Specification and Verification.} The related ideas in the specification of DNNs~\cite{propertyInference, seshia2018formal, usman2021nnrepair}.~\cite{seshia2018formal} discusses formalizing and reasoning about properties of DNN;  however,~\cite{seshia2018formal} does not propose any precondition inference using model architecture and post condition.~\cite{propertyInference} proposed a technique to compute input and layer properties from a feed-forward network and utilize 
formal contracts for the network. The application of inferred properties has been demonstrated to explain predictions,   guarantee robustness, simplify proofs, and network distillation.
~\cite{usman2021nnrepair} introduced a constraint-based technique for repairing neural network classifiers by inferring correctness specifications.~\cite{dreossi2019verifai} proposes a technique to apply formal methods to ML components e.g., perception systems, and analyze system behavior in an uncertain environment. However, ~\cite{propertyInference, usman2021nnrepair, dreossi2019verifai} did not consider abstracting neural networks and introduce a technique for computing data preconditions from trained DNN models and utilizing those inferred preconditions for implying trust in the model’s prediction during the deployment stage.
There is a recent study~\cite{shriverICSE21} on reducing DNN properties to enable falsification with adversarial attacks using a correctness problem comprised of a DNN and robustness problems property. 
In a recent study~\cite{rulemdfse21}, a rule induction-based technique has been proposed to facilitate the debugging process of trained statistical models only that generates an interpretable characterization of the data on which the predictive machine learning model performs poorly.
In another study~\cite{featureBias}, a bias-guided misprediction explanation technique has been proposed that generates explanation rules with higher misprediction explanation and also improves the machine learning model’s robustness utilizing a mispredicted area upweight sampling algorithm.
Recently, an empirical study~\cite{Khairunnesa2023} characterizes different kinds of ML contracts, which may help ML API developers to write contracts. Another research study~\cite{ahmed23dlcontract} proposed a technique for checking contracts for
deep learning libraries by specifying DL APIs with preconditions
and postconditions.
None of these recent papers along with the work~\cite{wang2023data, yang2022data} related to neural network specification and verification 
utilizes a DNN model's model architecture and expected output to infer assumptions on data that our approach emphasizes. We demonstrate the utility of inferred data preconditions to imply the trustworthiness in predicting unseen data during deployment. 
\section{Threats to Validity}
\label{sec:threats}
In the context of inferring preconditions from a deep learning model, internal threats to validity include an incorrect model structure where the DNN model may not fully capture the underlying system's complexity or dynamics, leading to inaccurate precondition inference.
External threats to validity 
include lack of representativeness in the unseen data where the data used to evaluate the model may not accurately reflect the real-world scenario, leading to the inaccurate implication of the model's prediction by our approach. 
To mitigate these threats, we have collected a large and diverse dataset that accurately represents the real-world scenario. This can help ensure the model is exposed to various variations and can generalize well to unseen data. Also, we have used more complex models with more \dense layers, which have the ability to learn complex patterns and features in the real-world dataset. 
\section{Conclusion and Future Work}
\label{sec:conclusion}
We propose a novel technique, \deepinfer, for inferring data preconditions from a DNN. 
\deepinfer uses an abstract representation of the DNN model and derived \wpre rules
for different types of DNN functions,
by solving challenges of non-linear computation with different dimensions of matrices,
to infer preconditions for the model. 
A DNN can be deployed with these preconditions, and their violation can imply trust 
in the model's predictions during deployment. 
We have evaluated \deepinfer on 29 models using 4 real-world datasets and found substantial results compared to prior work regarding effectiveness and efficiency. We find that data
precondition violations and incorrect model prediction are
highly correlated. \deepinfer effectively implies the correct and incorrect prediction of higher accuracy models with recall (0.98) and F-1 score (0.84), which is a significant improvement compared to prior work. \deepinfer is 3.29 times faster than the state-of-the-art technique.
In future, our approach can be extended to automatically validate the temporal properties of DNN models. We can also explore the use of predicate abstraction and symbolic reasoning for DNN models to further explain the black-box DNN models.
Recent studies on decomposing DNN into modules~\cite{decompose1, decompose2, decompose3}, we intend to infer input preconditions of each DNN module for its expected and reliable behavior. We want to extend our data precondition inference technique to mitigate model's unfairness ~\cite{sumon1, sumon2, usman} in different stages of the ML pipeline~\cite{sumon3}. We can enhance techniques~\cite{nguyen22manas, nguyen23fix} by inferring preconditions from mined models, considering improved accuracy for trustworthy prediction.
\section{Data Availability}
\label{sec:datapackage}
The replication packages and results are available in this repository~\cite{DIrepo} 
that can be leveraged by software engineering for machine learning research in the future.
\begin{acks}
We acknowledge the reviewers for their insightful
comments. This material is based upon work supported by the National Science Foundation under Grant CCF-15-18897, CNS-15-13263, CNS-21-20448, CCF-19-34884, and CCF-22-23812. All opinions are of the authors and do not reflect the view of sponsors.
\end{acks}
\balance
\bibliographystyle{ACM-Reference-Format}
\bibliography{deepInfer.bib}

\end{document}